\documentclass[reprint,bibnotes,amsmath,amssymb,aps,prb,showpacs,floatfix,superscriptaddress,longbibliography]{revtex4-1}

\usepackage{amsmath}
\usepackage{amsfonts}
\usepackage{amssymb}
\usepackage{amsxtra}
\usepackage{xcolor}
\usepackage{graphicx}
\usepackage{subfigure}
\usepackage{dcolumn}
\usepackage{float}
\usepackage{bm}
\usepackage[breaklinks=true,colorlinks,citecolor=blue,linkcolor=blue,urlcolor=blue]{hyperref}

\newcommand{\nn}{\nonumber}

\DeclareMathAlphabet{\bi}{OML}{cmm}{b}{it}
\def\be{\begin{equation}}
\def\ee{\end{equation}}
\def\bearr{\begin{eqnarray}}
\def\eearr{\end{eqnarray}}

\def\bs{\boldsymbol}

\begin{document}
\title{Nonlinear and anisotropic polarization rotation in two dimensional Dirac materials}
\author{Ashutosh Singh}
\affiliation{Department of Physics, Indian Institute of Technology Kanpur, Kanpur - 208016, India}
\author{Saikat Ghosh}
\affiliation{Department of Physics, Indian Institute of Technology Kanpur, Kanpur - 208016, India}
\author{Amit  Agarwal}
\email{amitag@iitk.ac.in}
\affiliation{Department of Physics, Indian Institute of Technology Kanpur, Kanpur - 208016, India}

\date{\today}

\begin{abstract}
We predict nonlinear optical polarization rotation in two dimensional massless Dirac systems including graphene and 8-$Pmmn$ borophene. When illuminated, a continuous wave optical field leads to a nonlinear steady state of photo-excited carriers in the medium. The photo-excited population inversion and the inter-band coherence gives rise to a finite transverse optical conductivity, $\sigma_{xy} (\omega)$. 
This in turn leads to definitive signatures in associated Kerr and Faraday polarization rotation, which are measurable in a realistic experimental scenario.
\end{abstract}

\maketitle
\section{Introduction}
Two dimensional (2D) electronic systems have garnered incredible attention over the past decade due to their exceptional opto-electronic properties and 
gate tunable response \cite{Bonaccorso,TMD_review}. Starting with graphene\cite{Novoselov, Geim}, several other 2D materials \cite{Butler}, including silicene~\cite{Feng,Ruge}, MoS$_2$~\cite{Mak} and phosphorene~\cite{Liu}, have been added to the list, each with its own peculiar regime of response. 
A comparatively recent addition to the 2D family of graphene, is one of the many 2D polymorphs of monolayer boron, termed as 8-$Pmmn$ borophene\cite {Mannix, Peng, Lopez, Feng1, Zabolotskiy}. Unlike graphene which has an isotropic electronic dispersion, 8-$Pmmn$ borophene has been shown to host a tilted and anisotropic Dirac cone\cite {Mannix, Peng, Lopez, Feng1, Zabolotskiy}. Such anisotropic and tilted Dirac cone dispersion has been shown to give rise to anisotropy in plasmon dispersion and screening \cite{Sadhukhan}, an extra contribution to the Weiss oscillations in the longitudinal magneto-conductivity \cite{PhysRevB.96.235405} and anisotropic optical conductivity in a linear response regime\cite{Verma}. Beyond linear response, one therefore expects such materials to manifest further richness in properties and accordingly, there has been an intense search for techniques to characterize such nonlinearities and use in device applications \cite{Davidovikj,doi:10.1021/acsnano.5b06110,doi:10.1021/acs.nanolett.7b04114}.

Here we explore nonlinear polarization rotation of a continuous wave (cw) optical field, reflected or transmitted from materials hosting quasiparticles with  a tilted and anisotropic Dirac cone dispersion. Usual studies related to optical properties in such materials\cite{Gusynin,Falkovsky,Mikhailov,Stauber,Wright,Ishikawa,Hendry,PhysRevLett.101.196405} focus 
on the linear response regime, ignoring higher order field induced changes in the photo-excited carrier distribution. However, with increasing optical field strength, the photo-excited carrier distribution reaches a nonlinear steady state through competing  rates of carrier excitation and various decay channels including carrier-carrier, carrier-phonon and impurity scattering \cite{Mishchenko,Singh,Chaves,Singh2}. Though such nonlinearities have been studied in a transient regime \cite{Dremetsika:16,2018arXiv180109785S}, a comparatively simpler steady-state study of the nonlinear transmission and reflection coefficients due to cw illumination, remains largely unexplored.

\begin{figure}[t!]
\begin{center}
\includegraphics[width=0.84 \linewidth]{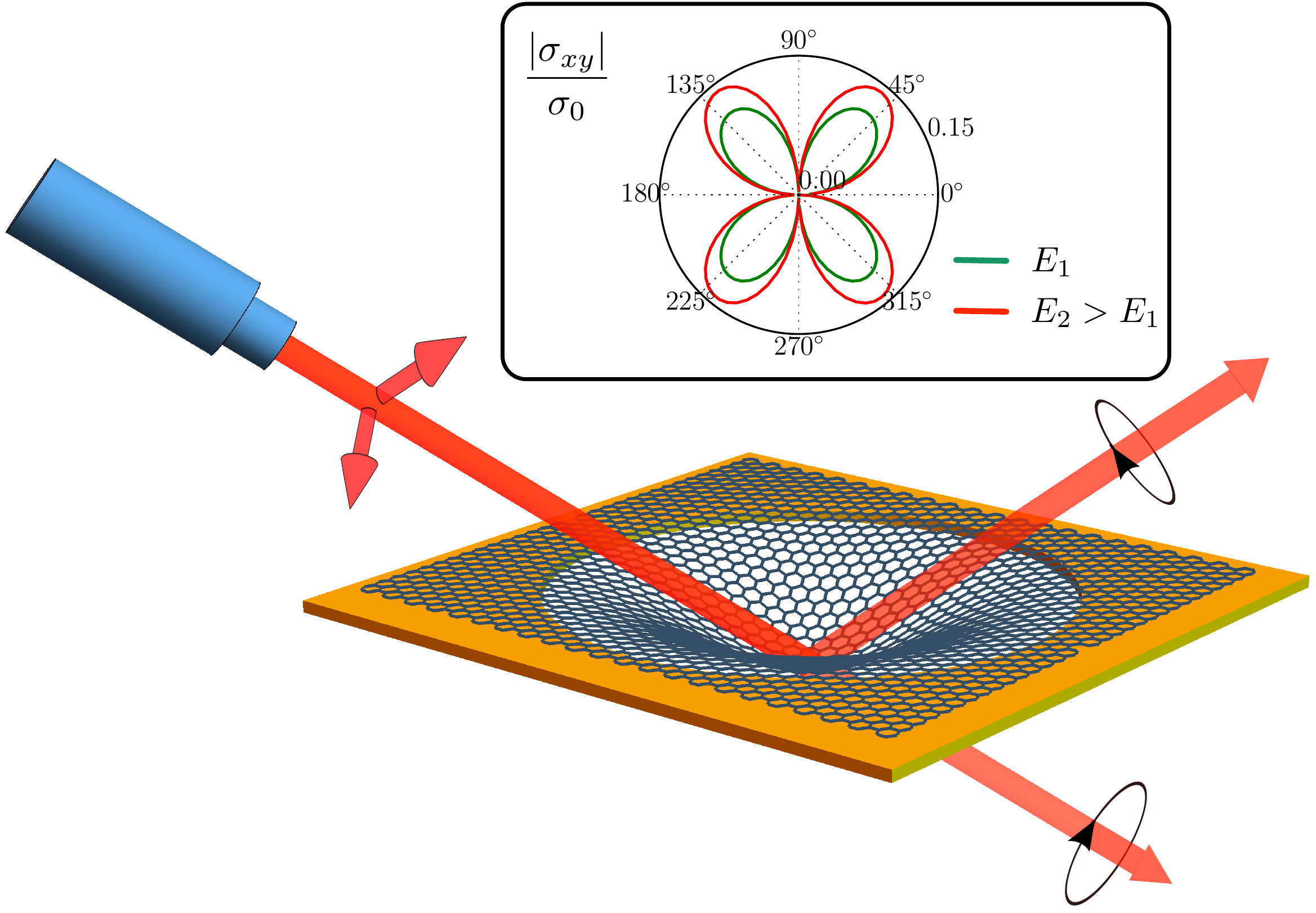}
\caption{A cartoon depicting a cw optical field undergoing polarization rotation in reflection and transmission when incident on a suspended graphene sheet. The polarization rotation occurs due to the emergence of a transverse optical conductivity ($\sigma_{xy}$), that bears a unique signature in the incident polarization angle ($\theta$) and the input field strength ($E_1$ and $E_2$, with $E_2> E_1$) as shown in the inset.}
 \label{Fig0}
 \end{center}
 \end{figure} 

We show emergence of a finite transverse optical conductivity $\sigma_{xy}(\omega)$, due to the nonlinear steady state population inversion, in a cw illuminated 2D  massless Dirac system [see inset of Fig.~\ref{Fig0}]. We estimate the corresponding polarization rotation of reflected and transmitted cw fields and predict a measurement regime that is accessible with reasonable optical field intensities. For experimental detection, the common mode linear response contribution to the polarization rotation from other sources, can be subtracted out by using a differential measurement technique for two differing intensities [see Fig.~\ref{Fig0}]. The predicted field controlled polarization rotation can open up new possibilities in nano-structured opto-electronic devices including fast polarization switches and dynamically controlled wave plates. 

This article is organized as follows: we first present the effective low energy Hamiltonian of 8-$Pmmn$ borophene hosting a tilted and anisotropic Dirac cone and use it to calculate the corresponding optical matrix elements in Sec.~\ref{HD}. With optical Bloch equations ~\cite{boyd2003nonlinear}, next we calculate the nonlinear steady state distribution function and the corresponding density matrix in presence of a cw field. 
The calculation of optical conductivity, transmission and reflection coefficients along with Kerr and Faraday rotation, is presented in Sec.~\ref{AOR} followed by a discussion of experimental implications in Sec.~\ref{S5}. 
Finally we summarize our results in Sec. \ref{Con}.
\section{Hamiltonian and the optical matrix element}\label{HD}
Hamiltonian for an electron interacting with an electric field is described in the dipole approximation as \cite{Aversa},
\be \label{eq1}
\hat H = \hat H_0 + e{\bf E}\cdot{\hat{\bf r}},
\ee
where $\hat H_0$ is the bare effective low energy Hamiltonian describing the the energy dispersion of the material, $e$ is the magnitude of the electronic charge, and ${\bf E}$ is the electric field vector.
The generalized low energy effective Hamiltonian for tilted and anisotropic gapless Dirac systems in 2D, in vicinity of the Dirac point is given by, 
$H = \sum_{\bf k} H_{\bf k}$, where 
\be \label{EqHam}
H_{\bf k} = \hbar(v_x\sigma_xk_x + v_y\sigma_yk_y + v_tk_y\mathbb{I}_{2\times2})~.
\ee
For the case of 8-$Pmmn$ borophene monolayer \cite{Zabolotskiy,Sadhukhan}, we have $v_x = 0.86 v_F$ and $v_y = 0.69 v_F$ as the carrier velocities in the $\hat x$ and $\hat y$ directions. 
 $v_t = 0.32 v_F$ is the tilt velocity and we have chosen $v_F = 10^6 ms^{-1}$. 
Here $\sigma_{x/y}$ represent the $x/y$ components of the three Pauli matrices and $\mathbb{I}_{2\times2}$ is a $2\times$2 unit matrix. For graphene, 
we generally have $v_t =0$, and $v_x = v_y = v_F$. However, in case of freestanding graphene which is clamped along the edges, i.e., in strained graphene we can have an anisotropic dispersion 
with $v_x \neq v_y$. 

The eigenvalues for $H_{\bf k}$ are given by,
\be\label{eigen}
\varepsilon_{\bf k}^{\lambda} = \hbar v_F{|\bf k|}\left[\tilde v_t\sin\phi_{\bf k} + \lambda( \tilde v_x^2\cos^2\phi_{\bf k} +  \tilde v_y^2\sin^2\phi_{\bf k})^{1/2}\right],
\ee
with ${|\bf k|} = (k_x^2 + k_y^2)^{1/2}$, and $\phi_{\bf k} = \tan^{-1}(k_y/k_x)$, the azimuthal angle and $\lambda$ takes values $+1$ and $-1$ for the conduction and valence band,  
respectively. 
In Eq.\eqref{eigen} we have defined the dimensionless velocities, 
$\tilde v_x = v_x/v_F$, $\tilde v_y = v_y/v_F$ and $\tilde v_t = v_t/v_F$. 

The dipole matrix element ${\bf r}^{\lambda\lambda'}~(\lambda\neq\lambda')$, in the basis formed by the eigenvectors of $H_{\bf k}$,  can be written in terms of momentum
matrix element (also called the optical matrix element) as~\cite{Singh}, ${\bf r}^{vc} = i{\bf M}_{\bf k}^{vc}/(e \omega_{\bf k})$, where we have defined $\hbar\omega_{\bf k} = \varepsilon_{\bf k}^{c} - \varepsilon_{\bf k}^{v}$.

For $H_{\bf k}$ of Eq.~\eqref{EqHam}, we have 
\be
{\bf M}_{\bf k}^{vc} = \frac{iev_F\tilde v_x\tilde v_y}{( \tilde v_x^2\cos^2\phi_{\bf k} +  \tilde v_y^2\sin^2\phi_{\bf k})^{1/2}}(\sin\phi_{\bf k},-\cos\phi_{\bf k}).
\ee
For the specific case of $v_x = v_y = v_F$, as in isotropic graphene,  we have ${\bf M}_{\bf k}^{vc} = iev_F(\sin\phi_{\bf k},-\cos\phi_{\bf k}).$

\section {Steady state photo-excited carriers and coherence}\label{SSPC}
In this section we consider optical pumping of 2D gapless Dirac system (monolayer 8-$Pmmn$ borophene or graphene) with a monochromatic continuous wave~(CW) laser. 
The field at any time $t$ is given as, ${\bf E} = E_0\cos\omega t~\hat{\bf e}$, with $E_0$ being the electric field amplitude, $\omega$ the optical laser frequency and 
$\hat{\bf e}$ the polarization direction. 

The dynamics of the system is best described in terms of the equation of  motion for the density matrix: 
$i \hbar \partial_t {\hat\rho}(t) = [H, \hat\rho]$. 
We denote the diagonal elements of the density matrix as $\rho_{11} = \rho^c_{\bf k}$ and $\rho_{22} = \rho^v_{\bf k}$, where $\rho_{\bf k}^\lambda \equiv \langle {a_{\bf k}^{\lambda}}
^\dagger a_{\bf k}^{\lambda} \rangle$ denotes the momentum resolved electron density in the $\lambda$-th band,  and the off diagonal elements of the 
density matrix as $\rho_{21} = p_{\bf k} \equiv \langle {a_{\bf k}^{c}}^\dagger a_{\bf k}^v \rangle$, and $\rho_{12} =  p_{\bf k}^*$. 
Here $p_{\bf k}$ is usually referred to as the inter-band coherence or polarization. Using these, and Eq.~\eqref{eq1}, the evolution of the density matrix 
is given by a set of coupled optical-Bloch equations \cite{Mishchenko,Singh,Chaves,Singh2}, 
\bearr \label{COBE3a}
\partial_t {n}_{\bf k} & =& 4 \Im m\left[ {\Omega}^{vc*}_{\bf k} {p}_{\bf k} \right] - \gamma_{1}({n}_{\bf k}-n^{\rm eq}_{\bf k})~, \\ 
\label{COBE3b}
\partial_{t}{p}_{\bf k} & = & i \omega_{\bf k}{p}_{\bf k}  - i {\Omega}^{vc}_{\bf k} {n}_{\bf k}-\gamma_{2}{p}_{\bf k}~.
\eearr
In Eq.~\eqref{COBE3a}-\eqref{COBE3b}, ${n}_{\bf k} \equiv \rho^c_{\bf k} - \rho^v_{\bf k}$ is generally referred to as population inversion and $\hbar {\Omega}^{vc} = e{\bf E}\cdot{\bf r}^{vc}$ is the interband Rabi frequency.
\begin{figure}[t!]
\begin{center}$
\begin{array}{cc}
\includegraphics[width = 0.99 \linewidth]{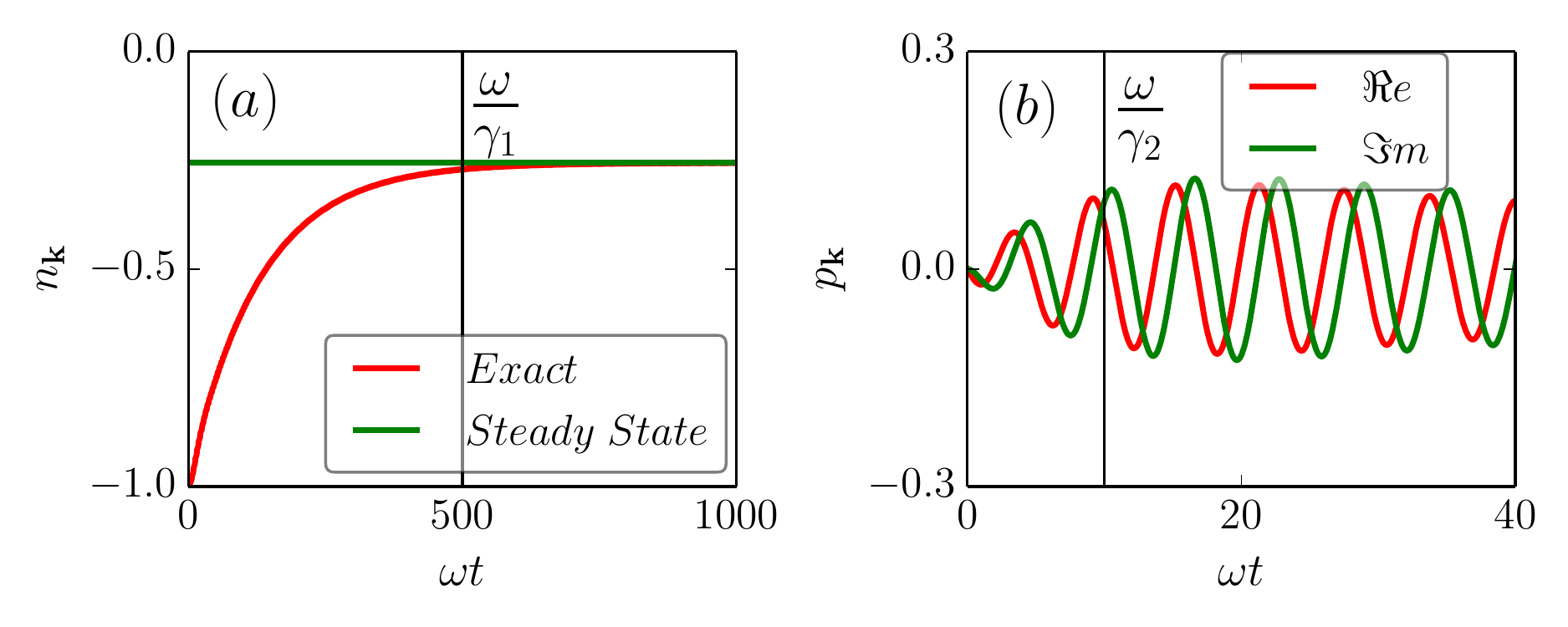}
\end{array}$
\caption{Time evolution of (a) $n_{\bf k}$~(shown in red) as a function of time (in units of $1/\omega$) along with its steady state value~(shown in green). Clearly, the long term dynamics of $n_{\bf k}$ as described by the exact
solution of Eqs. \eqref{COBE3a}-\eqref{COBE3b} increases from zero and saturates to its steady state value for $t\gg \gamma_1^{-1}$. (b) Exact time evolution of real and imaginary parts of microscopic polarization $p_{\bf k}$.
Here we have considered $(k_x,k_y) = (0.5,0.5)$, in units of $\omega/v_F$, $\zeta = 5$, $\omega = 5\times10^{14}s^{-1}$, $\gamma_1 = 10^{12}s^{-1}$, $\gamma_2 = 5\times10^{13}s^{-1}$, chemical potential $\mu = 0$ and the linear polarization angle $\theta = 0$.} 
\label{Fig1}
\end{center}
\end{figure}

The last two terms in both Eqs.~\eqref{COBE3a}-\eqref{COBE3b} are added phenomenologically to include the decay of the inverted population and the inter-band coherence \cite{Singh}. 
The equilibrium population inversion (in absence of light) is expressed as  $n^{\rm eq}_{\bf k} = f_{c\bf k}^{(0)} - f_{v\bf k}^{(0)}$, where $f_{\lambda{\bf k}}^{(0)}$  
denotes the Fermi function with band index $\lambda$, 
and $\gamma_{1}$, $\gamma_{2}$ are the phenomenological 
relaxation rate for the momentum resolved population inversion and  the inter-band coherence, respectively. Aiming for an insightful 
analytical solution, and for simplicity we assume $\gamma_1$ and $\gamma_2$ to be constants.

Solving Eqs.~\eqref{COBE3a}-\eqref{COBE3b} in the steady state, with the assumption\cite{Chaves} that,
$p_{\bf k} = p_{1\bf k}e^{i\omega t} + p_{2\bf k}e^{-i\omega t}$, we have,
\bearr\label{p1p2}
p_{1\bf k} &=& \frac{in_{\bf k}}{2\hbar\omega_{\bf k}}\frac{-{\bf E}\cdot{\bf M}_{\bf k}^{vc}}{\omega - \omega_{\bf k} - i\gamma_2}~, \nn \\ 
p_{2\bf k} &=& \frac{in_{\bf k}}{2\hbar\omega_{\bf k}}\frac{{\bf E}^*\cdot{\bf M}_{\bf k}^{vc}}{\omega + \omega_{\bf k} + i\gamma_2}~.
\eearr
Substituting  $p_{1\bf k}$ and $p_{2\bf k}$ from Eq. \eqref{p1p2} in Eq. \eqref{COBE3a}, in the steady state regime for which $\langle\partial_t n_{\bf k}\rangle_t = 0$, 
we obtain the following momentum resolved steady state value of 
the nonlinear distribution function (NDF) of the carrier population in the conduction and valence band, 
\bearr\nn\label{NDF}
\rho^{c}_{\bf k} \equiv f_{c \bf k}^{(1)} &=& \frac{1}{2}\left[f_{c \bf k}^{(0)}(1 + G_{\bf k}) + f_{v \bf k}^{(0)}(1 - G_{\bf k})\right],\\
\rho^{v}_{\bf k} \equiv f_{v \bf k}^{(1)} &=& \frac{1}{2}\left[f_{c \bf k}^{(0)}(1 - G_{\bf k}) + f_{v \bf k}^{(0)}(1 + G_{\bf k})\right].
\eearr
The nonlinearity and the anisotropy of the NDF is dictated by the function $G_{\bf k}$, which is 
explicitly given by \cite{Singh2},
\be \label{G}
 G_{\bf k} = \left[ 1 + \zeta^2~\frac{\omega^2}{\omega_{\bf k}^2} ~ \frac{2\gamma_2^2|\tilde {\bf M}^{vc}\cdot
 \hat{\bf e}|^2(\omega^2 + \omega_{\bf k}^2 + \gamma_2^2)}
 {\left[(\omega_{\bf k}^2 - \omega^2)^2 + 2\gamma_2^2(\omega_{\bf k}^2 + \omega^2) + \gamma_2^4\right]}\right]^{-1}.
\ee
Here, $\tilde {\bf M}^{vc} = {\bf M}^{vc}/(ev_F)$ is the dimensionless material dependent optical matrix element.
The optical field strength is embedded in the dimensionless parameter $\zeta$ in Eq.~\eqref{G}, which is given by 
\be\label{zeta}
\zeta \equiv \frac{e E_0 v_F}{\hbar \omega \sqrt{\gamma_{1} \gamma_{2}}}~. 
\ee
The parameter $\zeta$ is the optical field strength and frequency dependent main parameter, 
which characterizes the `degree' of the nonlinear effects and response. 

\begin{figure}[t!]
\begin{center}$
\begin{array}{cc}
\includegraphics[width=0.99 \linewidth]{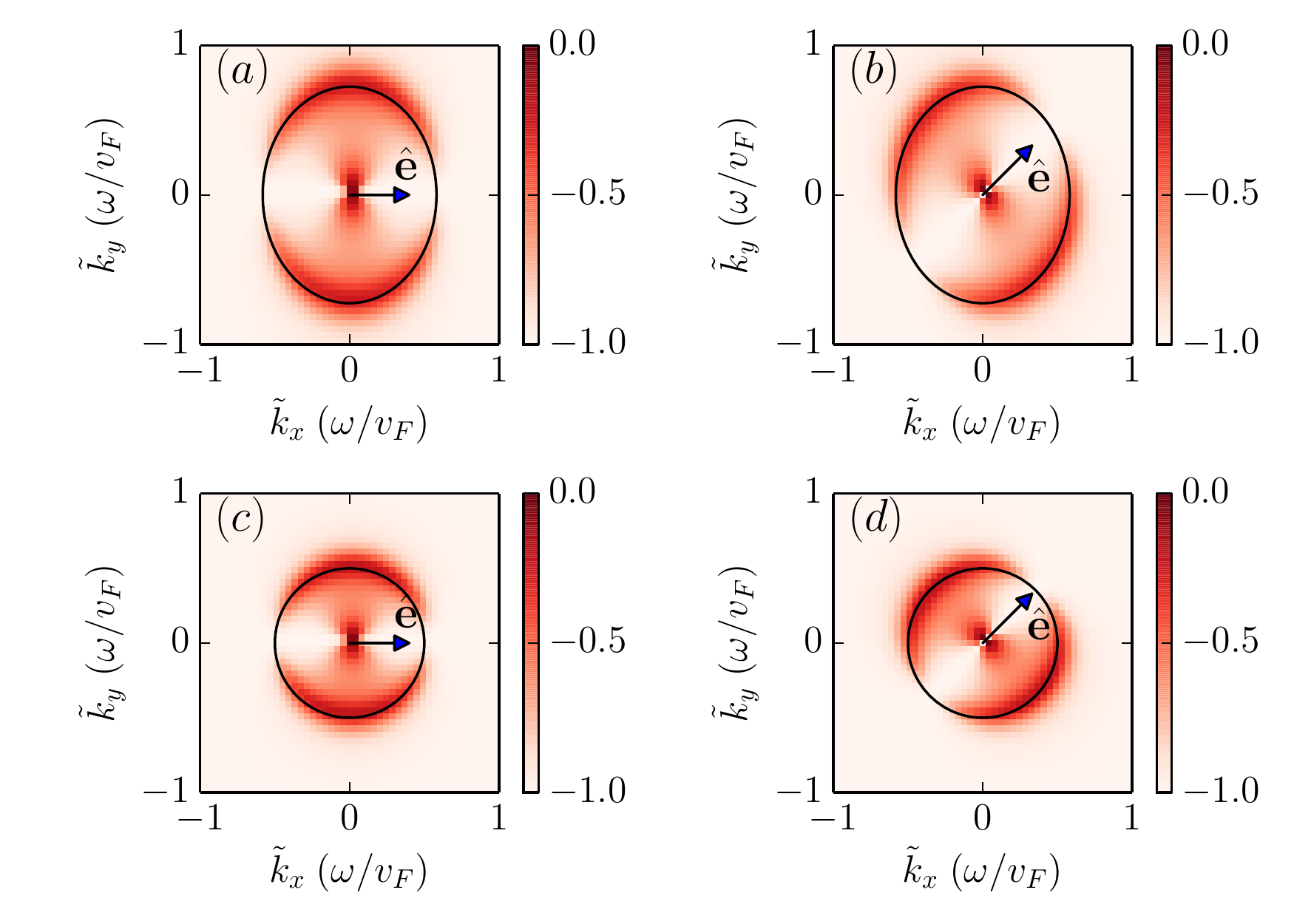}
\end{array}$
\caption{The steady state photo-excited population inversion function $n_{\bf k}$, in the $k_x-k_y$ plane, close to the Dirac point in 8-$Pmmn$ borophene [panels (a) and (b)] and graphene [panels (c) and (d)]for two different polarization direction indicated by $\hat {\bf e}$:  $\theta = 0$ [(a) and (c)] and $\theta = \pi/4$ [(b) and (d)]. 
The photo-excited population inversion is minimum along the polarization direction while it is maximum in the perpendicular direction. The solid black line marks the contour of $\omega_{\bf k} = \omega$, the phase space region around which the photo-excited carriers are centered. 
Here $\mu = 0$ and the other parameters are same as that of Fig.~\ref{Fig1}. } 
\label{Fig2}
\end{center}
\end{figure}

It is easy to check that in the absence of incident light beam, we have $\zeta\to0$, which implies $G_{\bf k}\to1$ and consequently $f_{\lambda \bf k}^{(1)}\to f_{\lambda \bf k}^{(0)}$, with no nonlinearity in the carrier distribution function. 
On the other hand, for very high intensity beams we have $\zeta\to\infty$, which leads to $G_{\bf k}\to0$ 
and this describes the optical saturation limit. Figure~\ref{Fig1} shows the time evolution of population inversion $n_{\bf k}$ and $p_{\bf k}$ as a function of time (in units of $1/\omega$). 
At $t = 0$ the carrier excitation probability is vanishingly small (with $n_{\bf k} = -1$), since the optical field is absent. With increasing time and after a few optical cycle, $n_{\bf k}$ starts increasing
and finally saturates to the steady state population inversion given by $n_{\bf k}^{\rm eq}G_{\bf k}$ over timescales given by $t \approx 1/\gamma_1 $. 

In the linear response regime, the optical conductivity is calculated using the Kubo formula. In the Kubo formula, the optical field generated nonlinearity in the carrier distribution function is neglected, i.e., 
$n_{\bf k} \to n^{\rm eq}_{\bf k}$. Also for convenience, the calculations are usually done in the infinite coherence time limit, i.e. $\gamma_2 \ll \omega$. To highlight the deviation in the carrier distribution function 
due to nonlinear optical effects in the same high frequency regime of $\omega \gg \gamma_2$, we expand $G_{\bf k}$ upto second order in $\zeta$ to yield  
\be\label{zeta_sq}
G^{\zeta^2}_{\bf k}\approx 1 - \pi\gamma_2\zeta^2~\frac{\omega^2}{\omega_{\bf k}^2}~|{\bf M}_{\bf k}^{vc}\cdot{\hat{\bf e}}|^2\delta(\omega_{\bf k}-\omega).
\ee
The factor $\gamma_2\zeta^2$ in the above expression is actually independent of $\gamma_2$ as $\zeta^2 \propto 1/\gamma_2$. 
This clearly highlights the fact that in the vanishingly small optical field limit, the carrier distribution function is primarily unchanged. Deviation from the equilibrium distribution of carriers, is captured by the term proportional to $\zeta^2$, which in energy is centered around 
$\omega_{\bf k} = \omega$. The $\zeta^2$ term is proportional to $|{\bf M}_{\bf k}^{vc}\cdot{\hat{\bf e}}|$, and this highlights the anisotropy in the NDF which arises from the anisotropic optical matrix element and the polarization direction. 

The highly anisotropic nature of the momentum resolved population inversion, $n_{\bf k}$, for 8-$Pmmn$ borophene is also highlighted in Fig.~\ref{Fig2}. Similar to the case of graphene \cite{Singh2}, the photo-excited carrier distribution has a maxima in a direction perpendicular to the direction of the optical field polarization. However unlike graphene, in borophene the profile of $n_{\bf k}$ is slightly elongated along the $y$-direction, and this can be attributed to the anisotropic band-structure of 8-$Pmmn$ borophene. This steady state photo-excited NDF, is what leads to nonlinear and anisotropic optical response in optical conductivity \cite{Mishchenko,Singh}, can lead to non-equilibrium plasmons \cite{Chaves}, giant and anisotropic photoconductivity \cite{Singh2}, or in the polarization rotation of the reflected and transmitted beam, as discussed in the current paper. 

\section{Nonlinear optical conductivity and Kerr rotation}\label{AOR}
\subsection{Steady state Optical conductivity}
Here we focus on the steady state nonlinear optical conductivity. 
Following our previous work\cite{Singh}, we can express the real part of the momentum resolved current density at any time $t$ in terms of microscopic polarization $p_{\bf k}$ and the optical matrix element ${\bf M}^{cv}_{\bf k}$
as,
\be \label{eq:J0}
{\bf J}_{\bf k}(t) =  -2 \Re e [p_{\bf k}(t) {\bf M}^{cv}_{\bf k}]~.
\ee
Thus the total current is given by,
\be\label{int_J}
{\bf J}(t) = \frac{g_sg_v}{4\pi^2}\int {\bf J}_{\bf k}(t)d{\bf k},
\ee
where $g_s$ and $g_v$ represents the spin and valley degeneracy respectively. The current in Eq.~\eqref{int_J} is real and it has terms with time dependence of $e^{i\omega t}$ as well as $e^{-i \omega t}$. In the spirit of the linear response theory, we use the $e^{-i \omega t}$ part of the current to define the real and the imaginary components of the optical conductivity (see details of the calculation in 
Eq.~\ref{current_fourier1} of the appendix). 
%
\begin{figure}[t!]
\begin{center}
\includegraphics[width=0.99 \linewidth]{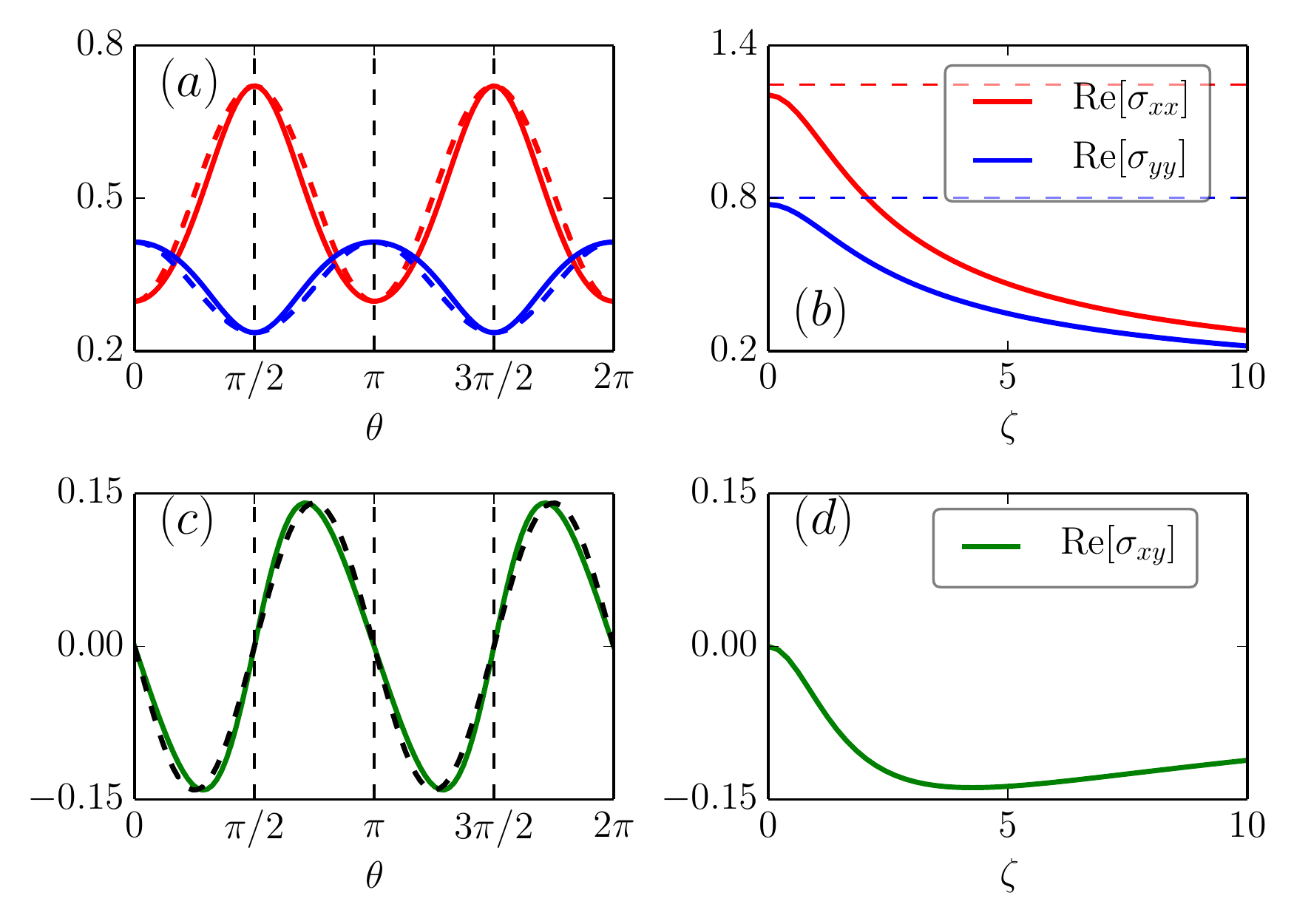}
\caption{The dependence of the interband optical conductivity of 8-$Pmmn$ borophene (in units of $\sigma_0~= e^2/{4\hbar}$), 
on the polarization angle ($\theta$) and 
the laser intensity (proportional to $\zeta^2$).
The Variation of the real part of $\sigma_{xx}$ and $\sigma_{yy}$ with $\theta$ for $\zeta = 5$ is shown in (a), whereas 
the variation of real part of $\sigma_{xy}$ is shown in (c). The exact angular dependence more or less follows the approximate angular dependence obtained for $\Re e(\sigma_{xx})$, $\Re e(\sigma_{yy})$ 
and $\Re e(\sigma_{xy})$ in Eqs.~(\ref{sigmaxxr_zeta2}), (\ref{sigmayyr_zeta2}) and (\ref{sigmaxyr_zeta2}), respectively. 
The nonlinear effects has been highlighted using the exact results in panel (b) and (d), which show the $\zeta$ dependence of the longitudinal and transverse optical conductivities, respectively, 
for $\theta = \pi/4$. Here $\mu = 0.2 \times \hbar \omega$ and the other parameters are same as that of Fig.~\ref{Fig1}.}
 \label{Fig3}
 \end{center}
 \end{figure} 
The nonlinear optical conductivities in  2D can be easily calculated via the relations, 
\be\label{sigma}
\sigma_{ij}(\omega) = \frac{g_s g_v}{(2\pi)^2}\int_{\rm BZ} d{\bf k}~ J_{i\bf k}(\omega)/E_j~,
\ee
where the integral is over the Brillouin zone of the material involved. 

While $\sigma_{ij}(\omega)$ is not difficult to calculate numerically, some useful insights can be obtained from the analytical expression retaining the 
first nonlinear term beyond the linear response regime. This is the term proportional to $\zeta^2$ or alternately to $E_0^2$, 
and depends on the intensity of the incident optical beam. The calculation simplifies in the  $\gamma_2 \ll \omega$ limit with the lorentzian in Eq.~\eqref{X_1} 
giving way to a delta function. In this limiting case, the real part of $x$ component of the longitudinal optical conductivity is given by,
\bearr \label{integrand} \nn
\frac{\Re e(\sigma_{xx}^{\zeta^2})}{\sigma_0} &=& \Theta(\hbar \omega - 2 \mu)\int \frac{d\phi_{\bf k}}{\pi}~\frac{\sin^2{\phi_{\bf k}} \tilde v_x^2 \tilde v_y^2}{( \tilde v_x^2\cos^2{\phi_{\bf k}} + 
 \tilde v_y^2\sin^2{\phi_{\bf k}})^2}\\
&\times&\left(1 - \frac{\zeta^2 \tilde v_x^2 \tilde v_y^2\sin^2({\phi_{\bf k} - \theta})}{ \tilde v_x^2\cos^2{\phi_{\bf k}} +  \tilde v_y^2\sin^2{\phi_{\bf k}}}\right).
\eearr
Here $\sigma_0~= e^2/{4\hbar}$ is the well known universal optical conductivity of graphene and $\theta$ denotes the polarization angle in an anticlockwise sense, with respect to the 
$\hat x$-axis of the crystal. 

Evidently, for $\hbar \omega < 2 \mu$ there are no vertical optical transitions possible due to Pauli blocking. Consequently all optical conductivities vanish for $\hbar \omega < 2 \mu$. 
Thus in the rest of the manuscript, we only discuss the case of $\hbar\omega > 2\mu$.

Performing the $\phi_{\bf k}$ integration for $\hbar\omega > 2\mu$ in Eq.~\eqref{integrand}, yields,
\be\label{sigmaxxr_zeta2}
\frac{\Re e(\sigma_{xx}^{\zeta^2})}{\sigma_0} = \frac{\tilde v_x}{8\tilde v_y}\left[8-\zeta^2(3 \tilde v_x^2 +  \tilde v_y^2)-\zeta^2(3 \tilde v_x^2 -  \tilde v_y^2)\cos2\theta\right].
\ee
The imaginary part of $\sigma_{xx}$ is obtained by using Eq.~\eqref{zeta_sq} in Eq.~\eqref{Jx_im}, which leads to,
\bearr\label{sigmaxxi_zeta2}
 \frac{\Im m(\sigma_{xx}^{\zeta^2})}{\sigma_0} &=& -\frac{1}{2\pi}~\frac{v_x}{v_y}~\ln\left[\frac{\gamma_2^2 + (\omega + 2\mu/\hbar)^2}{\gamma_2^2 +  (\omega - 2\mu/\hbar)^2}\right] \\ \nn
 &-&  \frac{\zeta^2}{16} \frac{\gamma_2}{\omega}~\frac{v_x}{v_y} \left[3v_x^2 + v_y^2 + (3v_x^2 - v_y^2)\cos2\theta\right].
\eearr
As an independent check of our formalism, the longitudinal optical conductivity of graphene ($v_x = v_y = v_F$) in the linear response regime of $\zeta\to 0$ limit is given by 
\be
\frac{\sigma_{xx}^{\zeta\to0}}{\sigma_0} = \Theta(\hbar \omega - 2 \mu) - \frac{i}{2\pi}~\ln\left[\frac{(\omega + 2\mu/\hbar)^2}{(\omega - 2\mu/\hbar)^2}\right]~, 
\ee
consistent with the results of Ref.~[\onlinecite{Falkovsky}]. 

Similarly the real and imaginary parts of the $y$ component of the longitudinal optical conductivity can also be calculated. They are explicitly given by 
\be\label{sigmayyr_zeta2}
\frac{\Re e(\sigma_{yy}^{\zeta^2})}{\sigma_0} = \frac{\tilde v_y}{8\tilde v_x}\left[8-\zeta^2(\tilde v_x^2 +  3\tilde v_y^2)-\zeta^2(\tilde v_x^2 - 3 \tilde v_y^2)\cos2\theta\right],
\ee
and,
\bearr\label{sigmayyi_zeta2}
 \frac{\Im m(\sigma_{yy}^{\zeta^2})}{\sigma_0} &=&-\frac{1}{2\pi}~\frac{v_y}{v_x}~\ln\left[\frac{\gamma_2^2 + (\omega + 2\mu/\hbar)^2}{\gamma_2^2 +  (\omega - 2\mu/\hbar)^2}\right] \\\nn
 &-& \frac{\zeta^2}{16}~\frac{\gamma_2}{\omega}~\frac{v_y}{v_x} \left[v_x^2 + 3v_y^2 + (v_x^2 - 3v_y^2)\cos2\theta\right].
\eearr 
Note that the anisotropy of the Dirac band-structure changes the universal isotropic optical conductivity of graphene (within linear response)
to  $\Re e(\sigma_{xx})\to \sigma_0 v_x/v_y$ and $\Re e(\sigma_{yy}) \to \sigma_0 v_y/v_x$. 

The transverse optical conductivity $\sigma_{xy}^{\zeta^2}$ can be  
obtained by replacing $\sin^2{\phi_{\bf k}}\to-\sin\phi_{\bf k}\cos\phi_{\bf k}$ in the numerator of the integrand in Eq.~\eqref{integrand}, and it is given by 
\be\label{sigmaxyr_zeta2}
\frac{\Re e(\sigma_{xy}^{\zeta^2})}{\sigma_0} = \frac{\Re e(\sigma_{yx}^{\zeta^2})}{\sigma_0} = -\frac{\zeta^2}{2}~\tilde v_x\tilde v_y\sin\theta\cos\theta,
\ee
and,
\be\label{sigmaxyi_zeta2}
\frac{\Im m(\sigma_{xy}^{\zeta^2})}{\sigma_0} = \frac{\Im m(\sigma_{yx}^{\zeta^2})}{\sigma_0} = -\frac{\zeta^2}{4}~\frac{\gamma_2}{\omega}~\tilde v_x\tilde v_y\sin\theta\cos\theta.
\ee
We emphasize that the transverse optical conductivity ($\sigma_{xy}$) vanishes in the linear response regime, and it is finite only when the nonlinear response (proportional to $\zeta$) is 
significant.

Numerically calculated exact real part of longitudinal~($\sigma_{xx}$, $\sigma_{yy}$) and transverse interband optical conductivity, for 8-$Pmmn$ borophene, 
as a function of $\theta$ and $\zeta$ are shown in Fig.~\ref{Fig3}. Panels (a) and (c) of Fig.~\ref{Fig3} have $\zeta =5$ and  finite $\mu = 10^{14}\hbar/s = \hbar \omega/5$ and show the polarization angle dependence
of the optical conductivities. Note that the exact numerical results are very close to the $\theta$ dependence predicted by approximate Eqs.~\eqref{sigmaxxr_zeta2}, \eqref{sigmayyr_zeta2},
and \eqref{sigmaxyr_zeta2}, 
i.e.,  $\Re e(\sigma_{xx}) \propto \cos 2\theta$,  
$\Re e(\sigma_{yy}) \propto \cos 2\theta$ and $\Re e(\sigma_{xy}) \propto \sin 2\theta$. This is highlighted by the dashed curves  
\footnote{The $\cos(2 \theta)$ dependent terms are fitted by an expression of the form, $y_1(\theta) = y_1^{\rm max} \sin^2\theta + y_1^{\rm min} \cos^2\theta $. On the other hand, the $\sin(2\theta)$ dependent terms are fitted by a form $y_2(\theta) = [y_2^{\rm max}-y_2^{\rm min}]\sin\theta \cos\theta $.} 
in Fig.~\ref{Fig3}.
In panels (b) and (d) of Fig.~\ref{Fig3} we fix $\theta =\pi/4$ and show the $\zeta$ dependence of the optical conductivity. For vanishingly small
intensity, $\zeta\to 0$, or in the linear response regime,  we have $ n_{\bf k}\to n^{\rm eq}_{\bf k}$ and therefore $\Re e(\sigma_{xx})\to \sigma_0 v_x/v_y = 1.25 \sigma_0$ and 
$\Re e(\sigma_{yy}) \to \sigma_0 v_y/v_x = 0.8 \sigma_0$ and $\Re e(\sigma_{xy}) \to 0$. 

\begin{figure}[t!]
\begin{center}$
\begin{array}{cc}
\includegraphics[width=0.95 \linewidth]{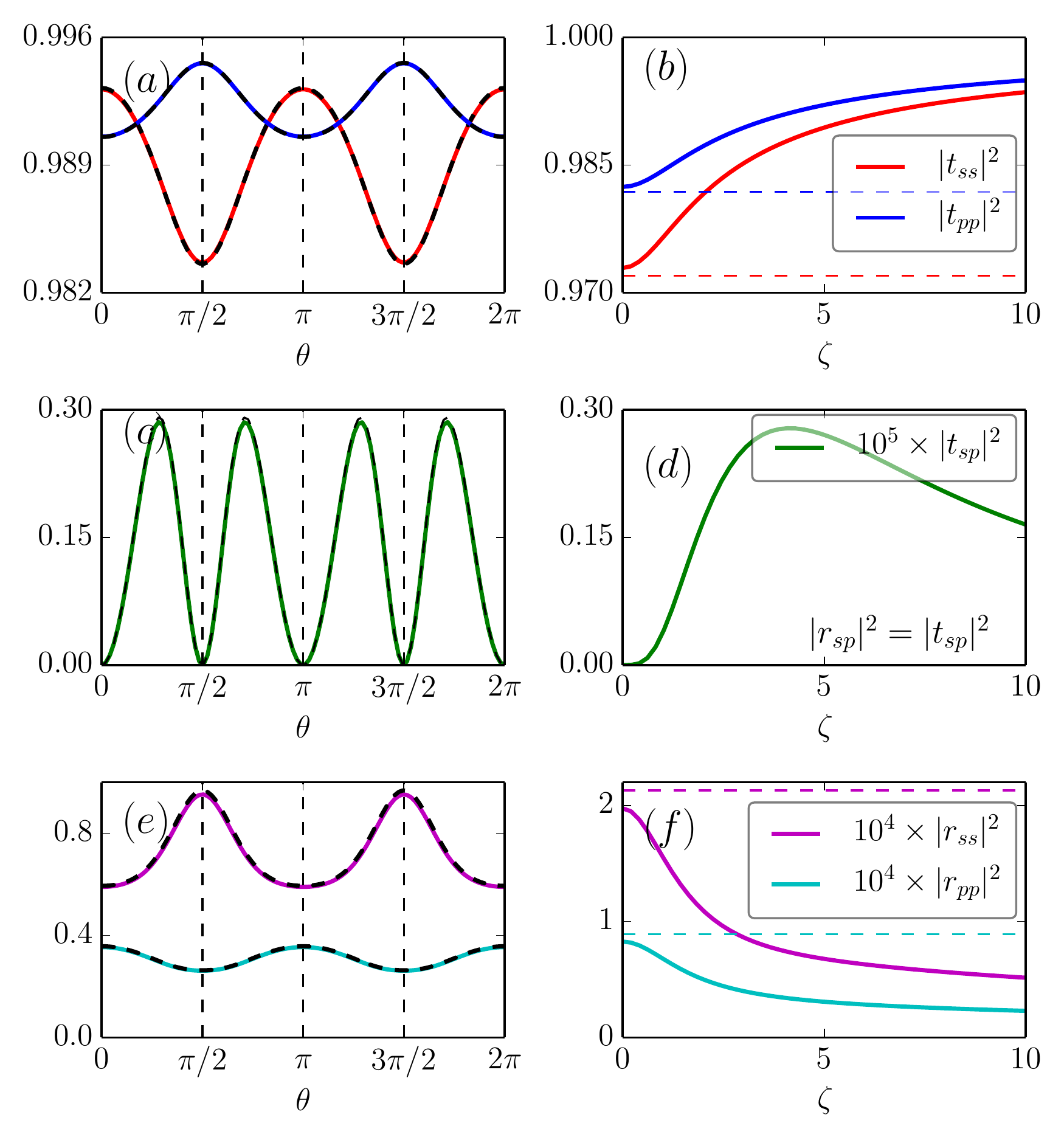}
\end{array}$
\caption{Polarization and intensity dependence of different components of the transmission and reflection probability. (a) The variation of diagonal transmission probability for $s-$ and $p-$ components, $|t_{ss}|^2$ and $|t_{pp}|^2$ with $\theta$ for $\zeta = 5$. The off diagonal transmission/reflection probability is shown in (c).  The $\theta$ dependence of the two diagonal components of the reflection probability are shown in (e). 
The approximate form of their angular variation (marked with dashed black curve) in (a), (c) and (e) are given by Eqs.~\eqref{tss_alpha}, \eqref{tpp_alpha} and \eqref{rsp_alpha}. In panel (b) (d) and (e) we have shown the $\zeta$ dependence of the different components of the transmission and reflection probabilities for $\theta = \pi/4$. All parameters here are identical to that of Fig.~\ref{Fig3}.
}
\label{Fig4}
\end{center}
\end{figure}

\subsection{Optical transmission and reflection}
Physically, optical conductivity sets the reflected, transmitted or absorbed optical spectrum along with rotation of the 
polarization angle in either the reflected (Kerr rotation) or in the transmitted (Faraday rotation) optical beam. 
Accordingly, we next derive a general expression for transmission and reflection coefficients for both $s-$ and $p-$ polarized incident optical beam  
and explore their implications for 2D massless Dirac materials, with particular emphasis on graphene and 8-$Pmmn$ borophene. For simplicity we assume the graphene or the borophene monolayer to be perfectly 2D (without ripples and defects), a reasonable assumption for 
an optical beam with diffraction spot size. 

The transmission and reflection coefficients for $s-$ (in the plane of incidence) and $p-$ (perpendicular to the plane of incidence) polarized optical beam can be expressed in terms of the 
complex optical conductivity of the 2D monolayer \cite{Yoshino:13}. For the sake of completeness, we reproduce the calculations 
in Appendix \ref{A1}. Using the fact that the optical conductivities in 2D materials are typically of the order of $\sigma_0 = e^2/(4 \hbar)$, the exact expressions for the 
transmission and reflection coefficients can be simplified. 

Retaining only the first order terms in the small parameter, $\pi\alpha_F/2$ where $\alpha_F \sim 1/137$ is the fine structure constant, the transmission and reflection coefficients can be expressed as Eqs.~\eqref{tss_alpha}-\eqref{rsp_alpha}. Thus the transmission coefficient is simply given by 
\be
|t_{ss}|^2 \approx \left(1 - \frac{\pi\alpha_F}{2}\frac{\Re e(\sigma_{xx})}{\sigma_0}\right)^2 +  \frac{\pi^2\alpha_F^2}{4}\frac{\Im m(\sigma_{xx})^2}{\sigma_0^2}.
\ee
Although both real and imaginary parts of the optical conductivity contribute to the transmittance, we can safely ignore the imaginary part for all practical purpose.
In fact, keeping only first order term in $\alpha_F$ suffices and we have, $|t_{ss}|^2 \approx 1 - \pi\alpha_F~\Re e(\sigma_{xx})/\sigma_0$. Similarly, the diagonal component of the 
transmittance for the $p-$ component is given as, $|t_{pp}|^2 \approx 1 - \pi\alpha_F~\Re e(\sigma_{yy})/\sigma_0$. Therefore, for an anisotropic system, where $\sigma_{xx}\neq\sigma_{yy}$,
we have,
\be
|t_{pp}|^2 - |t_{ss}|^2 \approx \frac{\pi \alpha_F}{\sigma_0}\left[{\Re e(\sigma_{xx})} - \Re e(\sigma_{yy})\right].
\ee

For a 2D system like graphene which has an isotropic band-structure in vicinity of the Dirac point with $v_x = v_y = v_F$, we have 
 $\sigma_{xx} = \sigma_{yy} =  \sigma_0$ and hence, $|t_{ss}|^2_{\zeta\to 0} = |t_{pp}|^2_{\zeta\to 0} \approx 1-\pi\alpha_F$ \cite{Falkovsky,Nair1308}.
However for a system with anisotropic massless Dirac band-structure as in 8-$Pmmn$ borophene, we have 
$|t_{pp}|^2_{\zeta\to 0} \approx 1-\pi\alpha_F v_y/v_x$, while $|t_{ss}|^2_{\zeta\to 0} \approx 1-\pi\alpha_F v_x/v_y$.
Thus in 8-$Pmmn$ borophene the transmittance for  $s-$ and $p-$ polarized optical beam are not identical, even in the linear response regime. 
This transmission anisotropy can in fact be used to measure the ratio of the anisotropic Dirac velocities, via the relation 
\be
|t_{pp}|^2_{\zeta\to 0} - |t_{ss}|^2_{\zeta\to 0} \approx \pi\alpha_F\left(\frac{v_x}{v_y}-\frac{v_y}{v_x}\right).
\ee

Numerically exact components of transmission and reflection probability are presented in Fig.~\ref{Fig4}. Panels (a), (c) and (e) show their  
dependence on the polarization angle $\theta$ for 8-$Pmmn$ borophene.  We have chosen $\zeta = 5$, $\theta_i = \theta_t = 0$ and $n_i = n_t = 1$. 
Similar to the optical conductivity, here also we see that the exact results match reasonably well with the $\theta$ dependence predicted by the approximate relations, 
Eqs.~\eqref{tss_alpha}, \eqref{tpp_alpha} and \eqref{rsp_alpha} as indicated by the dashed lines in panels (a), (c) and (e) of Fig.~\ref{Fig4}.  
The impact of varying optical field strength is shown in panels (b), (d) and (f). 

\subsection{Polarization rotation}
We next use these derived reflection and transmission coefficients to calculate Kerr and Faraday polarization rotation angle. In essence, the existence of finite off diagonal components of the coefficients, $\{r_{sp},r_{ps},t_{sp},t_{ps}\}$, which in turn arise due to finite $\sigma_{xy}(\omega)$, leads 
polarization rotation. Such polarization rotation in the reflected beam is measurable and has been experimentally explored in graphene, in presence of a 
perpendicular magnetic field\cite{Shimano,Crassee} leading to a finite $\sigma_{xy}$. 
In the present context, the origin of finite $\sigma_{xy}$ is essentially the nonlinear response in the optical field strength \cite{Vasko}. We find that while the polarization rotation is estimated to be small, it is within measurable regime of existing experimental techniques.

\begin{figure}[t!]
\begin{center}$
\begin{array}{cc}
\includegraphics[width=0.95 \linewidth]{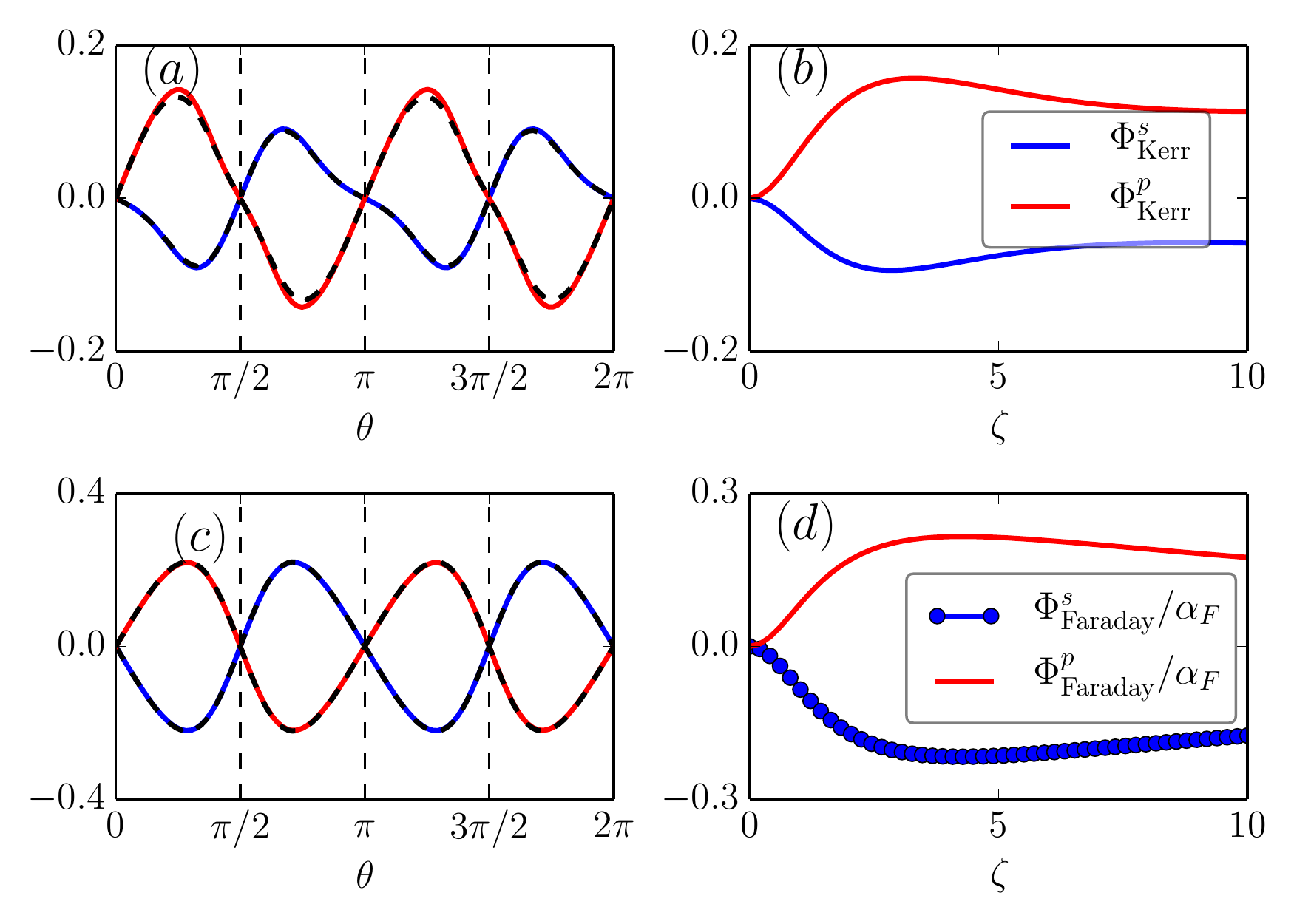}
\end{array}$
\caption{Kerr and Faraday rotation as a function of the polarization angle and the optical intensity of the incident beam. Panel (a) and (c) shows the exact $\Phi_{\rm Kerr}^{s/p}$ and $\Phi_{\rm Faraday}^{s/p}$ 
as a function of $\theta$ for $\zeta = 5$, 
respectively. The corresponding dashed black curves follow the angular dependence described by Eq.~\eqref{phi_Kerr} and Eq.~\eqref{phi_Faraday}. The intensity dependence of the corresponding polarization rotation angles is shown in (b) and (c) for $\theta = \pi/4$. Note that the $s-$ and $p-$ components of Kerr and Faraday rotation angles are out-of phase with respect to each other.  All parameters here are identical to that of Fig.~\ref{Fig3}.
}
 \label{Fig5}
\end{center}
\end{figure}

In general, the rotation in the polarization angle in the reflected beam can be expressed in terms of the following dimensionless complex quantities defined by  
\be\label{Kerr}
\chi_{\rm Kerr}^s = -\frac{r_{ps}}{r_{ss}}, ~~~~{\rm and}~~~~ \chi_{\rm Kerr}^p = \frac{r_{sp}}{r_{pp}}~, 
\ee
where the superscript $p$ and $s$ denote the $s-$ or $p-$ polarization of the incident beam. 
Similarly the rotation in the polarization angle in the transmitted beam can be expressed in terms of the following dimensionless complex numbers \cite{Yoshino:13},  
\be 
\label{Faraday}
\chi_{\rm Faraday}^s = -\frac{t_{ps}}{t_{ss}}, ~~~~{\rm and}~~~~ \chi_{\rm Faraday}^p = \frac{t_{sp}}{t_{pp}}.
\ee
Now the polarization angle $\Phi$, the azimuth of the major axes of the polarization ellipse of the reflected or transmitted beam, is given by \cite{Yoshino:13, 0953-8984-28-37-375802}
\be \label{Kerr2}
\tan(2 \Phi_{\rm M}^{s/p}) = \frac{2 {\rm Re}~[\chi_{\rm M}^{s/p}]}{1 - |\chi_{\rm M}^{s/p}|^2}~, 
\ee
where the subscript ${\rm M} = $ Kerr (Faraday) for the reflected (transmitted) optical beam. 
The ellipticity $\varepsilon$, or the major - minor axis ratio of the corresponding polarization ellipse 
is given by \cite{Yoshino:13}
\be
\label{Kerr3} 
\varepsilon_{\rm M}^{s/p} = \tan \left(\frac{1}{2} \sin^{-1} \left[\frac{2 {\rm Im [\chi_{\rm M}^{s/p}]}}{1 + |\chi_{\rm M}^{s/p}|^2} \right] \right).
\ee 
We emphasize that in Eqs.~\eqref{Kerr2}-\eqref{Kerr3}, it is essential to include the imaginary part of the optical transmission and reflection coefficients, 
for an accurate evaluation of the polarization rotation angle and the ellipticity \cite{Yoshino:13,0953-8984-28-37-375802,MacDonald,Nandkishore}. 
Note that in the limiting case of small optical fields, such that $\zeta^2 \ll 1$, the polarization angle dependence of both $\Phi_{\rm M}^{s/p}$ and $\varepsilon_{\rm M}^{s/p}$, 
is approximately given by $\sin(2\theta)$ [see Eqs.~\eqref{chi_s_Kerr}-\eqref{chi_p_Faraday}]. 
Both $\Phi_{\rm M}^{p/s}$, and $\varepsilon_{\rm M}^{p/s}$ can be experimentally measured by spectroscopic ellipsometry\cite{fujiwara2007spectroscopic}. 

\begin{figure}[t!]
\begin{center}$
\begin{array}{cc}
\includegraphics[width=0.95 \linewidth]{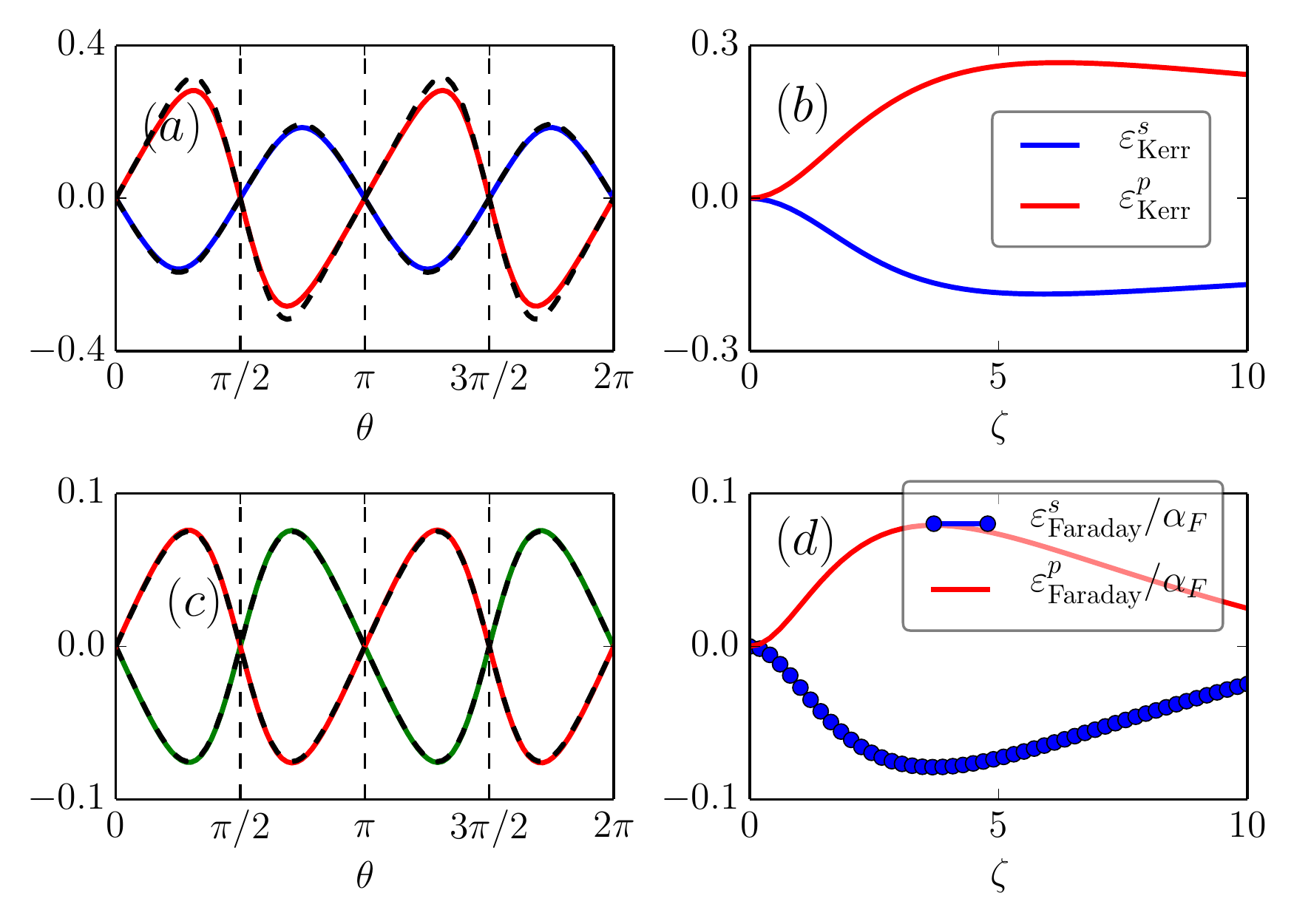}
\end{array}$
\caption{The ellipticity for the reflected and transmitted light, as a function of the polarization angle and optical intensity of the incident beam. Panel (a) and (c) shows the exact $\varepsilon_{\rm Kerr}^{s/p}$ and $\varepsilon_{\rm Faraday}^{s/p}$ as a function of 
 $\theta$ for $\zeta = 5$. The corresponding dashed black curves follow the angular dependence described by Eqs.~\eqref{phi_s_Kerr}-\eqref{phi_p_Kerr}. 
The intensity dependence of the two are shown in panels (b) and (c) 
for $\theta = \pi/4$ respectively.  All parameters here are identical to that of Fig.~\ref{Fig3}.
}
 \label{Fig6}
\end{center}
\end{figure}

The exact dependence of the polarization rotation angle on $\theta$ and $\zeta$ is shown in Fig.~~\ref{Fig5}. 
As expected, the angular dependence of the 
polarization rotation is reasonably captured by the approximate expressions in Eq. \eqref{phi_Kerr} and Eq. \eqref{phi_Faraday}. Similarly Fig.~\ref{Fig6}, 
shows the $\theta$ and $\zeta$ dependence of the ellipticity.  We propose that 
the $\theta$ dependence of the polarization rotation and the ellipticity as shown in Figs.~\ref{Fig5}-\ref{Fig6}, and the variation of their magnitude with changing optical field strength, can be 
used experimentally to study nonlinear optical effects in various 2D materials. 

\section{Experimental implications}\label{S5}
We finally explore viability of measuring the proposed nonlinear Kerr rotation in graphene or 8-$Pmmn$ borophene. 
High quality, single crystal free standing  monolayer graphene samples are routinely fabricated by several groups and furthermore, magneto-optical Kerr and Faraday rotation in graphene has been measured experimentally\cite{Shimano,Crassee}. Simultaneously, with precision polarization elements (Glan-Thomson polarizers, for example) a rotation accuracy of $10^{-6}$ radians is routinely achieved in optics experiments, setting the limit of measurement.

An estimate of the optical field strength needed to experimentally observe the polarization rotation can be obtained from Eq.~\eqref{chi_s_Kerr}. To zeroth order in $\alpha_F$, Eq.~\eqref{chi_s_Kerr} reduces to  
\be 
\label{eq:chi_s_approx}
\chi_{\rm Kerr}^s \approx - \zeta^2 \frac{\tilde{v}_y^2 \sin(2 \theta)}{4}~.
\ee
Accordingly, Eq.~\eqref{eq:chi_s_approx} sets a practical limit $\zeta^2 \sin{2\theta} > 4 \times 10^{-6}$ for graphene and we use this as the basis of our estimation. 
Furthermore, the limit translates to a requirement of $\zeta > 2 \times 10^{-3}$, or equivalently for the optical field strength to be 
$E_0 > 4\hbar \omega \sqrt{\gamma_1 \gamma_2}/(e v_F) \times 10^{-6}$. Assuming $\hbar \omega = 1.5$ eV  and $\sqrt{\gamma_1 \gamma_2} \approx 10^{13}$ s$^{-1}$, we would need the optical field to be of the order of $E_0 \ge 3 \times 10^{4}$ V/m, which is easily achievable with current technology.

In case of graphene or 8-$Pmmn$ borophene, the measured nonlinear Kerr angle 1) will vary in a manner similar to $\sin(2\theta)$ as a function of the polarization angle, and 2) depend on the strength of the optical field. 

Experimentally, however, there are other possible effects which can also lead to polarization rotation in suspended graphene sheets. These include the presence of impurities, static folds and ripples, gate induced bending of the graphene sheet and effects of clamping at the edge of the sample. 
Nevertheless, it can be noted that none of these effects can lead to, 1) a $\sin(2\theta)$ like dependence of the Kerr angle on the polarization angle and  2) the optical field strength dependence of the Kerr angle ($E_0^2$ for small fields). In fact, the resulting Kerr angle rotation due to all these stray effects is expected to be completely independent of the strength of optical field strength.
 
A simple experimental way to separate these effects from the proposed nonlinear response in the Kerr angle, is to measure the resulting Kerr rotation at a given location in the suspended graphene sheet as a function of $\theta$ for two different values of the optical field intensity.
All stray rotation signals should be common mode to both the measurements, thereby dropping off, when a difference is taken. The observation of the 
$\sin(2\theta)$ like dependence in the difference would be a definitive signature of observing and estimating optical nonlinearities of 2D materials with Dirac cone like dispersion. 

\section{Conclusion}\label{Con}
We have presented a study of nonlinear and anisotropic optical response of 2D massless Dirac materials like graphene and the 8-$Pmmn$ borophene. 
Starting from a two band Hamiltonian and considering the carrier-field interaction in length gauge, we have obtained analytical expressions for the nonlinear steady state density matrix elements i.e., steady state population inversion and inter-band coherence, of the photo-excited carriers.   
The photo-excited population inversion and the interband coherence lead to a finite transverse optical conductivity $\sigma_{xy}(\omega)$, beyond the linear response regime. 
This in turn modifies the transmission and reflection coefficients, giving rise to finite $\{r_{sp},r_{ps},t_{sp},t_{ps}\}$ bearing definitive signatures in Faraday and Kerr polarization rotations. 

In particular, the Kerr angle has a unique $\sin(2 \theta)$ like dependence on the linear polarization angle of the incident beam along with a strong dependence on the incident field strength. We conclude that the nonlinear polarization rotation for graphene and 8-$Pmmn$ borophene is measurable with current techniques and can lead to applications in designing polarization switches, field controlled polarization controllers and wave plates.
\appendix 
\section{The real and imaginary part of the optical conductivity}
Momentum resolved interband current for a generic two band Hamiltonian is defined as~\cite{Singh},
\be
{\bf J_k} = -2{\rm Re}\left[p_{\bf k}{\bf M}^{cv}_{\bf k} \right],
\ee
where $p_{\bf k}$ is the interband coherence and ${\bf M}^{cv}_{\bf k}$ is an off-diagonal component of optical transition matrix denoting a transition
from valence to conduction band. Since~\cite{Chaves, Singh2},
\be
p_{\bf k} = p_{1\bf k}e^{i\omega t} + p_{2\bf k}e^{-i\omega t},
\ee
we have,
\bearr 
{\bf J_k} &=& -p_{1\bf k}{\bf M}^{cv}_{\bf k}e^{i\omega t} - p_{2\bf k}{\bf M}^{cv}_{\bf k}e^{-i\omega t}\\
&-& p^*_{1\bf k}{\bf M}^{vc}_{\bf k}e^{-i\omega t} - p^*_{2\bf k}{\bf M}^{vc}_{\bf k}e^{i\omega t}. 
\eearr
To define the real and imaginary part of the current response, it is generally expressed as 
${\bf J_k}(t) = \frac{1}{2}\left({\bf J_k}(\omega)e^{-i\omega t} + H.c.\right)$. Thus we have
\be\label{current_fourier1}
\frac{1}{2}{\bf J_k}(\omega) = -p_{2\bf k}{\bf M}^{cv}_{\bf k} - p^*_{1\bf k}{\bf M}^{vc}_{\bf k},
\ee
or more explicitly, 
\be\label{current_fourier} 
{\bf J_k}(\omega) = -\frac{in_{\bf k}}{\hbar\omega_{\bf k}}
\left[\frac{\left({\bf E}\cdot{\bf M}^{vc}_{\bf k}\right){\bf M}^{cv}_{\bf k}}{(\omega + \omega_{\bf k} + i\gamma_2)} + 
\frac{\left({\bf E}\cdot{\bf M}^{cv}_{\bf k}\right){\bf M}^{vc}_{\bf k}}{(\omega - \omega_{\bf k} + i\gamma_2)}\right].
\ee

To separate the current response into its real and imaginary part, we note 
that in general the optical transition matrix can be a complex quantity. For 2D systems we can express it as, 
\be 
{\bf M}^{vc}_{\bf k} = \left(M^{x}_r + iM^{x}_i\right)~\hat {\bf x} + \left(M^{y}_r + iM^{y}_i\right)~\hat {\bf y},
\ee
and ${\bf M}^{vc}_{\bf k} = ({\bf M}^{cv}_{\bf k})^*$. 
Using this we have,
\bearr \label{A8}
\left[\left({\bf E}\cdot{\bf M}^{vc}_{\bf k}\right){\bf M}^{cv}_{\bf k}\right]_{x} &=& E_x|M^x|^2 + E_y\sum_{l = r,i}M^x_l M^y_l \nn \\
&+&iE_y\sum_{\substack {p,q = r,i\\ p\neq q}}\epsilon_{pq}M^x_p M^y_q~.
\eearr
Similarly, 
\bearr \label{A9}
\left[\left({\bf E}\cdot{\bf M}^{vc}_{\bf k}\right){\bf M}^{cv}_{\bf k}\right]_{y} &=& E_x\sum_{l = r,i}M^x_l M^y_l + E_y|M^y|^2  \nn \\
&-& i E_x \sum_{\substack {p,q = r,i\\ p\neq q}}\epsilon_{pq}M^x_p M^y_q,
\eearr
where we have defined $\epsilon_{ri} = 1$ and $\epsilon_{ir} = -1$.
The $x-$ and $y$ components of $\left({\bf E}\cdot{\bf M}^{cv}_{\bf k}\right){\bf M}^{vc}_{\bf k}$ can be found by taking complex conjugate of the above expressions.
Substituting Eqs.~\eqref{A8}-\eqref{A9} in Eq.~\eqref{current_fourier}, the real and imaginary parts of the current can be obtained to be 
\begin{widetext}
\bearr
{\rm Re}(J_x) &=& -\frac{n_{\bf k}}{\hbar\omega_{\bf k}}\left[\left(E_x|M^x|^2 + E_y\sum_{l = r,i}M^x_l M^y_l\right)X_1
+ E_y\sum_{\substack {\left\lbrace p,q = r,i\right\rbrace\\ p\neq q}}~\epsilon_{pq}M^x_p M^y_q~Y_2\right]\label{Jx_re},
\eearr
\bearr
{\rm Im}(J_x) &=& -\frac{n_{\bf k}}{\hbar\omega_{\bf k}}\left[-E_y\sum_{\substack {\left\lbrace p,q = r,i\right\rbrace\\ p\neq q}}~\epsilon_{pq}M^x_p M^y_q~X_2
+ \left(E_x|M^x|^2 + E_y\sum_{l = r,i}M^x_l M^y_l\right)Y_1\right]\label{Jx_im},
\eearr
\bearr
{\rm Re}(J_y) &=& -\frac{n_{\bf k}}{\hbar\omega_{\bf k}}\left[\left(E_x\sum_{l = r,i}M^x_l M^y_l + E_y|M^y|^2\right)X_1
- E_x\sum_{\substack {\left\lbrace p,q = r,i\right\rbrace\\ p\neq q}}~\epsilon_{pq}M^x_p M^y_q~Y_2\right]\label{Jy_re},
\eearr
\bearr
{\rm Im}(J_y) &=& -\frac{n_{\bf k}}{\hbar\omega_{\bf k}}\left[E_x\sum_{\substack {\left\lbrace p,q = r,i\right\rbrace\\ p\neq q}}~\epsilon_{pq}M^x_p M^y_q~X_2
+ \left(E_x\sum_{l = r,i}M^x_l M^y_l + E_y|M^y|^2\right)Y_1\right]\label{Jy_im}. \nn \\ 
\eearr
\end{widetext}
Here we have defined the following:
\be
X_1 =  \frac{\gamma_2}{(\omega-\omega_{\bf k})^2 +\gamma_2^2} + \frac{\gamma_2}{(\omega + \omega_{\bf k})^2 +\gamma_2^2}~,\label{X_1}
\ee
\be
Y_1 =  \frac{\omega-\omega_{\bf k}}{(\omega-\omega_{\bf k})^2 +\gamma_2^2} + \frac{\omega +\omega_{\bf k}}{(\omega + \omega_{\bf k})^2 +\gamma_2^2}~,\label{y_1}
\ee
\be
X_2 =  \frac{\gamma_2}{(\omega-\omega_{\bf k})^2 +\gamma_2^2} - \frac{\gamma_2}{(\omega + \omega_{\bf k})^2 +\gamma_2^2}~,\label{X_2}
\ee
\be
Y_2 =  \frac{\omega-\omega_{\bf k}}{(\omega-\omega_{\bf k})^2 +\gamma_2^2} - \frac{\omega +\omega_{\bf k}}{(\omega + \omega_{\bf k})^2 +\gamma_2^2}~.\label{Y_2}
\ee
Till now the formalism is very general and works for any 2D material. Below we discuss the implications for massless Dirac fermions with an anisotropic and tilted Dirac cone -
as in 8-$Pmmn$ Borophene. 

Expressing Eq.~\eqref{EqHam}. in terms of the generic two band Hamiltonian of Refs.~[\onlinecite{Singh,Singh2}], $\hat H = \sum_{\bf k}h_{\bf k}\cdot{\bs\sigma}$, we have 
$h_{0\bf k} = \hbar v_t k_y$, $h_{1\bf k} = \hbar v_x k_x$, $h_{2\bf k} = \hbar v_y k_y$, and $h_{3\bf k} = 0$.
The corresponding band dispersion is given by 
\be
\varepsilon^{\lambda}_{\bf k} =  \hbar v_t k_y + \lambda \sqrt{\hbar^2 v_x^2 k_x^2 + \hbar^2 v_y^2 k_y^2},
\ee
where $\lambda = -~(+)$ for valence~(conduction) band. The transition frequency is given by,
\be
\omega_{\bf k} = \varepsilon^{+}_{\bf k} - \varepsilon^{-}_{\bf k} =  2 v_F |{\bf k}| \sqrt{\tilde{v}_x^2 \cos^2\phi_{\bf k} + \tilde{v}_y^2 \sin^2\phi_{\bf k}}. 
\ee
The corresponding 2D phase space integration factor is 
\be
 |{\bf k}| d{\bf k} = \frac{\omega_{\bf k}d\omega_{\bf k} d \phi_{\bf k}}{ 4 v_F^2\left(\tilde{v}_x^2 \cos^2\phi_{\bf k} + \tilde{v}_y^2 \sin^2\phi_{\bf k}\right)}.
\ee
The corresponding optical matrix elements are given  by \cite{Singh, Singh2},
\be
{\bf M}^{vc}_{\bf k} = \frac{ie}{\hbar h_{\bf k}}(h_2\nabla_{\bf k}h_1 - h_1\nabla_{\bf k}h_2). 
\ee
For the case of Eq.~\eqref{EqHam}, we have 
\be
\nabla_{\bf k}h_1 = \left(\frac{\partial (\hbar v_x k_x)}{\partial k_x}, \frac{\partial (\hbar v_x k_x)}{\partial k_y}\right) = \left(\hbar v_x, 0\right),
\ee
\be
\nabla_{\bf k}h_2 = \left(\frac{\partial (\hbar v_y k_y)}{\partial k_x}, \frac{\partial (\hbar v_y k_y)}{\partial k_y}\right) = \left(0, \hbar v_y\right),
\ee
finally,
\be
{\bf M}^{vc}_{\bf k} = iev_F\tilde M_{\bf k}\left(\sin\phi_{\bf k}, -\cos\phi_{\bf k}\right),
\ee
with,
\be
\tilde M_{\bf k} = \frac{\tilde v_x \tilde v_y}{\sqrt{\tilde{v}_x^2 \cos^2\phi_{\bf k} + \tilde{v}_y^2 \sin^2\phi_{\bf k}}}.
\ee
The real part of the longitudinal conductivity, $\sigma_{xx}$ can be obtained from Eq. \ref{Jx_re}. In the regime of allowed optical transitions, i.e., 
$\hbar \omega > 2 \mu$, it is given by  
\bearr\label{sigma_xx}
{\rm Re}(\sigma_{xx}) &=& {\rm Re}(J_x)/E_x \\ 
&=& \frac{\sigma_0}{\pi^2}\int~\frac{d\omega_{\bf k}d\phi_{\bf k}~G_{\bf k}\tilde M^2\sin^2\phi_{\bf k}}{\left(\tilde{v}_x^2 \cos^2\phi_{\bf k} + \tilde{v}_y^2 
\sin^2\phi_{\bf k}\right)}
X_1~. \nn 
\eearr
The imaginary component can simply be obtained by replacing $X_1$ with $Y_1$ in the above expression. 
Note that to evaluate Eq.~\eqref{sigma_xx} numerically based on a continuum model, we have to use a band cutoff. 
In this paper we use the energy cutoff of 8~eV for all calculations,  based on half bandwidth of the tight-binding Hamiltonian of 
graphene.  
The calculation of all other conductivity components 
proceeds along similar lines. 

\section{The reflection and transmission coefficients in terms of conductivity}
\label{A1}
\begin{figure}[b]
\centering
\includegraphics[width=0.8\linewidth]{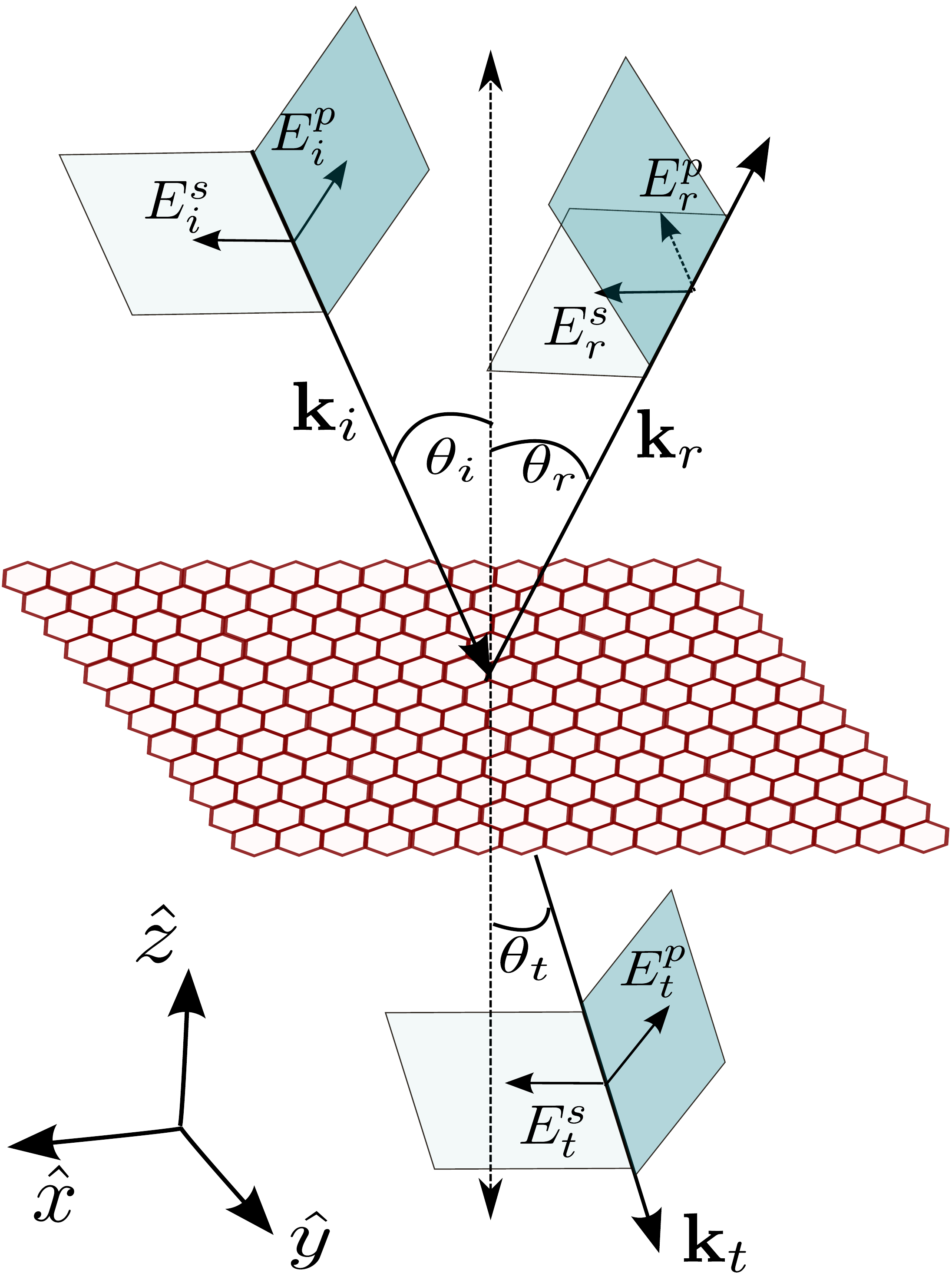}
\caption{Schematic for calculation of transmission and reflection coefficients. Incident light falls on the 2D material at an angle $\theta_i$ w.r.t. $\hat z$ axis
and is reflected and transmitted at $\theta_r$ and $\theta_t$ respectively. Both $s-$ and $p-$ components are shown for incident~(${\bf E}_i$), reflected
~(${\bf E}_r$) and the transmitted~(${\bf E}_t$) light along with their corresponding wave vectors ${\bf k}_i$, ${\bf k}_r$ and ${\bf k}_t$. 
The $p-$ polarization of the incident, reflected and transmitted beams is in the $y-z$ plane which is also the plane of incidence. 
The $s-$ polarization in all three beams is along the $x$ axis. 
The dotted perpendicular line coincides with the $\hat z$ axis.}
\label{Fig7}
\end{figure}
To start with we resolve the electric field vector of the incoming optical beam into $s$- and $p$- components and then we find the associated transmission and reflection coefficients. 
The existence of an off-diagonal ($sp$) component
in the transmission/reflection matrix is ensured by finite non-zero value of $\sigma_{xy}$ for an arbitrary polarization angle $\theta$. Following Ref.~[\onlinecite{peatross2011physics}],  
let ${\bf E}_i$, ${\bf E}_r$ and ${\bf E}_t$ be the electric field vector for
the incoming, reflected and the transmitted part of the light beam, respectively,  and ${\bf B}_i$, ${\bf B}_r$ and ${\bf B}_t$ be the corresponding magnetic field vectors. 
If the wave vectors and frequencies for the same are denoted as ${\bf k}_i, {\bf k}_r$, and ${\bf k}_t$, and $\omega_i, \omega_r$, and $\omega_t$ respectively, then
from Fig.~\ref{Fig7}, we have 
\bearr\nn
{\bf E}_i &=& (E_i^s, E_i^p\cos\theta_i, E_i^p\sin\theta_i)e^{({\bf k}_i\cdot{\bf r}-\omega_it)},\\\nn
{\bf E}_r &=& (E_r^s, -E_r^p\cos\theta_r, E_r^p\sin\theta_r)e^{{\bf k}_r\cdot{\bf r}-\omega_rt)},\\\nn
{\bf E}_t &=& (E_t^s, E_t^p\cos\theta_t, E_t^p\sin\theta_t)e^{({\bf k}_t\cdot{\bf r}-\omega_tt)}.\nn
\eearr
Now using ${\bf B} = (n/c)~{\bf\hat k}\times{\bf E}$, with $n$ denoting the refractive index of the medium, we have
\bearr\nn
{\bf B}_i &=& \frac{n_i}{c}(E_i^p, -E_i^s\cos\theta_i, -E_i^s\sin\theta_i)e^{({\bf k}_i\cdot{\bf r}-\omega_it)},\\\nn
{\bf B}_r &=& \frac{n_r}{c}(E_r^p, E_r^s\cos\theta_r, -E_r^s\sin\theta_r)e^{({\bf k}_r\cdot{\bf r}-\omega_rt)},\\\nn
{\bf B}_t &=& \frac{n_t}{c}(E_t^p, -E_t^s\cos\theta_t, -E_t^s\sin\theta_t)e^{({\bf k}_t\cdot{\bf r}-\omega_tt)}.\nn
\eearr
The fields at the interface of the monolayer with air (or vacuum)  satisfy the following boundary condition,
\be \label{eqE1}
 {\bf E}_1^{||} = {\bf E}_2^{||},~~~~{\rm and}~~~~~~~{\bf B}_1^{||}-{\bf B}_2^{||} = \mu_0{\bf J}\times{\bf{\hat z}}~,
\ee
where ${\bf E}_1 = {\bf E}_i + {\bf E}_r$, ${\bf E}_2 = {\bf E}_t$, ${\bf B}_1 = {\bf B}_i + {\bf B}_r$, ${\bf B}_2 = {\bf B}_t$
and ${\bf J}$ is the current density produced in the monolayer due to the incident optical beam. Here we will ignore the possibility of higher harmonics generation \cite{PhysRev.128.606}, 
which anyway happens for very high intensity laser beams. We also assume that the medium on either side of the monolayer is identical. 
The Snell's law  follows
from the first boundary condition. By matching the space and time dependent exponents:  $({\bf k}_i\cdot{\bf r}-\omega_it) = ({\bf k}_r\cdot{\bf r}-\omega_rt) =
({\bf k}_t\cdot{\bf r}-\omega_tt)$ for the monolayer located at $z =0$ we obtain  $\omega_i = \omega_r = \omega_t$, $\cos\theta_i = \cos\theta_r$, and $n_i\sin\theta_i = n_t\sin\theta_t$. 
Matching the $x$ and $y$ components of both the electric and magnetic fields at $z = 0$, as per Eq.~\eqref{eqE1}, we
arrive at a matrix equation:
\be \label{eq:matrix}
{\mathbb{S}}_{4 \times 4} \mathbb{E}_o = {\mathbb{E}_i}~. 
\ee
Here ${\mathbb E}_o = [E_t^s, E_r^s, E_t^p, E_r^p]^{T}$ represents the $s$ and $p$ components of the outgoing fields, 
\begin{widetext}
\be \label{eqS4}
{\mathbb{S}_{4\times4}} = 
\begin{pmatrix}
 1 & -1 & 0 & 0 \\
 0 & 0 & \cos\theta_t & \cos\theta_i\\
 \mu_0\sigma_{yx} & 0 & n_t/c + \mu_0\sigma_{yy}\cos\theta_t & -n_i/c\\
n_t\cos\theta_t/{c} + \mu_0\sigma_{xx}  & n_i\cos\theta_i/c & \mu_0\sigma_{xy}\cos\theta_t & 0\\
\end{pmatrix},
\ee
\end{widetext}
and ${\mathbb E}_i = [E_i^s, E_i^p\cos\theta_i, n_iE_i^p/c, n_i\cos\theta_iE_i^s/c]^T$ comprises of the incoming fields.
Solving the matrix Eq.~\eqref{eq:matrix}, the transmitted electric fields in the $p$- and $s$ directions can be expressed as 
\be
\begin{pmatrix}
 E_t^p\\
 E_t^s\\
\end{pmatrix} = \begin{pmatrix}
                t_{pp} & t_{ps}\\
                t_{sp} & t_{ss}\\
                \end{pmatrix} 
\begin{pmatrix}
 E_i^p \\
 E_i^s\\
 \end{pmatrix}.
\ee
Here the diagonal part of the transmission coefficients are given by,
\be\label{eq:tss}
t_{ss} = \left. \frac{E_t^s}{E_i^s}\right|_{E_i^p = 0} = \frac{2n_i}{c\mu_0}\frac{\sigma_2}{\sigma_T}\cos\theta_i,
\ee
and,
\be\label{eq:tpp}
t_{pp} = \left. \frac{E_t^p}{E_i^p}\right|_{E_i^s = 0} = \frac{2n_i}{c\mu_0}\frac{\sigma_1}{\sigma_T}\cos\theta_i~.
\ee
Here we have defined  
\bearr 
\sigma_T &=& \left(\sigma_1\sigma_2 - \sigma_{xy}\sigma_{yx}\cos\theta_i\cos\theta_t \right), 
\eearr 
along with 
\bearr
\sigma_1 &=& n_i\cos\theta_i/(c\mu_0) + n_t\cos\theta_t/(c\mu_0) + \sigma_{xx}~, \\
\sigma_2 &=& n_i\cos\theta_t/(c\mu_0) + n_t\cos\theta_i/(c\mu_0) + \sigma_{yy}\cos\theta_i\cos\theta_t~. \nn
\eearr
The off diagonal elemants of the transmission matrix are given by,
\be\label{eq:tsp}
t_{sp} = \left. \frac{E_t^s}{E_i^p}\right|_{E_i^s = 0} = -\frac{2n_i}{c\mu_0}\frac{\sigma_{xy}}{\sigma_T}\cos\theta_i\cos\theta_t,
\ee
and,
\be\label{eq:tps}
t_{ps} = \left. \frac{E_t^p}{E_i^s}\right|_{E_i^p = 0} = -\frac{2n_i}{c\mu_0}\frac{\sigma_{yx}}{\sigma_T}\cos^2\theta_i~.
\ee

Similar to the transmission coefficients, the reflection coefficients can also be estimated. Expressing the reflection coefficients in terms of the transmission coefficients, we have 
\bearr\label{rss}
r_{ss} && ~= t_{ss} - 1~,\\
\label{rps}
r_{ps} &&~ = -t_{sp} \frac{\sigma_{yx}}{\sigma_{xy}}~,\\
\label{rsp}
r_{ps} &&~ = -t_{ps}\frac{\cos\theta_t}{\cos\theta_i}~,~~~\text{and}~~~\\
\label{rpp}
r_{pp} &&~= 1 - t_{pp}\frac{\cos\theta_t}{\cos\theta_i}~.\\\nn
\eearr
\subsection{Lowest order ($\zeta^2$) nonlinear correction in the reflection and transmission coefficients}
Similar to the case of optical conductivity, while the exact numerical evaluation of the transmission and reflection coefficients is possible, it is insightful to express
the different reflection and transmission coefficients analytically upto second order in the optical field strength.
Using Eqs.~\eqref{sigmaxxr_zeta2}-\eqref{sigmaxyi_zeta2} in Eqs.~\eqref{eq:tss}-\eqref{eq:tps}, we derive the expressions for the  different transmission coefficients upto $\zeta^2$. 
The expression for $s-$ and $p-$ components of longitudinal transmission coefficients upto $\zeta^2$~(valid for $\zeta\ll1$) are,
\begin{widetext}
\be\label{t_ss_zeta2}
t_{ss}^{\zeta^2} = \frac{2}{\pi\alpha_F}\frac{\left(\alpha + \frac{2}{\pi\alpha_F}\right) + ik_3\alpha}{k_3^2\alpha^2 + \left(\alpha + \frac{2}{\pi\alpha_F}\right)^2} +
\frac{\zeta^2}{4\pi\alpha_F}\frac{\alpha(k_1 + k_2 \cos2\theta)}{\left[k_3^2\alpha^2 + \left(\alpha + \frac{2}{\pi\alpha_F}\right)^2\right]^2}
\left[\left(\alpha + \frac{2}{\pi\alpha_F}\right)^2 - k_3^2\alpha^2 + 2i\alpha\left(\alpha + \frac{2}{\pi\alpha_F}\right)k_3\right]~,
\ee
and,
\be\label{t_pp_zeta2}
 t_{pp}^{\zeta^2} = \frac{2}{\pi\alpha_F}\frac{\alpha\left(1 + \frac{2}{\pi\alpha_F}\alpha + ik_3\right)}{k_3^2 + \left(1 + \frac{2}{\pi\alpha_F}\alpha\right)^2} + \frac{\zeta^2}{4\pi\alpha_F}\frac{\alpha(k_4 + k_5 \cos2\theta)}{\left[\left(1 + \frac{2}{\pi\alpha_F}\alpha\right)^2 + k_3^2\right]^2}
\left[\left(1 + \frac{2}{\pi\alpha_F}\alpha\right)^2 - k_3^2 + 2i\left(1 + \frac{2}{\pi\alpha_F}\alpha\right)k_3\right]~.
\ee
\end{widetext}
Here we have defined, $k_1 = 3\tilde v_x^2 + \tilde v_y^2$, $k_2 = 3\tilde v_x^2 - \tilde v_y^2$, 
\be
k_3 = \frac{1}{2\pi} \ln\left[\frac{(\omega + 2\mu/\hbar)^2}{(\omega - 2\mu/\hbar)^2}\right]~,
\ee
$k_4 = \tilde v_x^2 + 3\tilde v_y^2$, $k_5 = \tilde v_x^2 - 3\tilde v_y^2$, and $\alpha = \tilde v_x/\tilde v_y$ .
Interestingly the presence of a finite $\sigma_{xy}(\omega)$, allows for the transmission of a s- (or p-) polarized beam as a p- (or s-) polarized beam. 
The complex off diagonal transmission coefficient is given by 
\begin{widetext}
\bearr\label{t_sp_zeta2}
t_{sp}^{\zeta^2} = \zeta^2\frac{\tilde v_x^2}{2\pi\alpha_F}\sin2\theta\frac{\Bigg[\left(\alpha + \frac{2}{\pi\alpha_F}\right)\left(1 + \alpha\frac{2}{\pi\alpha_F}\right)-k_3^2\alpha + 
ik_3\left(\alpha + \frac{2}{\pi\alpha_F} + \alpha\left(1 + \alpha\frac{2}{\pi\alpha_F}\right)\right)\Bigg]}{\left[k_3^2\alpha^2 + \left(\alpha + \frac{2}{\pi\alpha_F}\right)^2\right]\left[k_3^2 +
\left(1 + \frac{2}{\pi\alpha_F}\alpha\right)^2\right]}~. 
\eearr
\end{widetext}
Note that $t_{sp}^{\zeta^2} \propto \zeta^2$, and hence it is finite only due to the nonlinear optical response.  
Similar expressions for reflection coefficients can be obtained be using above expressions in Eqs. \eqref{rss}-\eqref{rpp}. 

Applying Eqs.~\eqref{Kerr}-\eqref{Faraday} to the case of anisotropic 2D gapless Dirac materials, 
we obtain the following expressions for s- and p- components of $\chi_{\rm Kerr}$ upto $\zeta^2$ to be,
\be\label{chi_s_Kerr}
\chi^s_{\rm Kerr} = \frac{\zeta^2\eta(\theta)\left[k_3^2 - \left(1 + \frac{2}{\pi\alpha_F}\alpha\right) -2ik_3\left(1+\alpha\frac{1}{\pi\alpha_F}\right)\right]}{\alpha\left[k_3^2 +
\left(1 + \frac{2}{\pi\alpha_F}\alpha\right)^2\right]}~,
\ee
and,
\be\label{chi_p_Kerr}
\chi^p_{\rm Kerr} = \frac{-\zeta^2\eta(\theta)\left[k_3^2\alpha - \left(\alpha + \frac{2}{\pi\alpha_F}\right) -2ik_3\left(\alpha + \frac{1}{\pi\alpha_F}\right)\right]}{k_3^2\alpha^2 + \left(\alpha + \frac{2}{\pi\alpha_F}\right)^2}.
\ee
Here we have defined $\eta(\theta) = \tilde v_x^2\sin2\theta/[2\pi\alpha_F\left(k_3^2 + 1\right)]$. 
Similarly the $s-$ and $p-$ components of $\chi_{\rm Faraday}$ upto $\zeta^2$ are given by,
\be\label{chi_s_Faraday}
\chi^s_{\rm Faraday} = -\zeta^2\frac{\eta(\theta)\pi\alpha_F}{2}\frac{(1 + k_3^2)\left[\left(1 + \frac{2}{\pi\alpha_F}\alpha\right) + ik_3\right]}{\left[k_3^2 + \left(1 + \frac{2}{\pi\alpha_F}\alpha\right)^2\right]}~,
\ee
and,
\be\label{chi_p_Faraday}
\chi^p_{\rm Faraday} = \zeta^2\frac{\eta(\theta)\pi\alpha_F}{2}\frac{(1 + k_3^2)\left[\left(\alpha + \frac{2}{\pi\alpha_F}\right) + ik_3 \alpha\right]}
{\alpha\left[k_3^2\alpha^2 + \left(\alpha + \frac{2}{\pi\alpha_F}\right)^2\right]}.
\ee
 All of these $\chi_{\rm M}^{p/s}$, show a $\sin(2 \theta)$ dependence on the angle of polarization of the incident beam. Thus 
in order to have non-vanishing polarization rotation, the polarization direction of the incoming beam should make
a finite angle with the $x$-axis, which denotes one of the principal crystal axis of the 2D gapless Dirac material.
\subsection{Simplified form of polarization rotation and ellipticity upto lowest order in the fine structure constant}
Here we specifically discuss the case of vertical incidence, i.e., $\theta_i = \theta_t = 0$. The case of generic 
incidence angle can be easily obtained from Eq.~\eqref{eq:tss}-\eqref{rpp}. %
For vertical incidence we have, 
$r_{sp} = t_{ps}$ and $r_{pp} = 1 - t_{pp}$. 
Now using the fact that $\sigma_{xy} = \sigma_{yx}$, and defining the normalized conductivity ${\tilde \sigma}_{ij} = \sigma_{ij}/\sigma_0$, 
we have 
\be\label{tss}
t_{ss} = \frac{2}{\pi\alpha_F}\frac{\left(2/{\pi\alpha_F} + {\tilde \sigma}_{yy}\right)}{\left(2/{\pi\alpha_F} + {\tilde \sigma}_{xx}\right)\left(2/{\pi\alpha_F} + {\tilde \sigma}_{yy}\right)
-{\tilde \sigma}_{xy}^2},
\ee
\be\label{tpp}
t_{pp} = \frac{2}{\pi\alpha_F}\frac{\left(2/{\pi\alpha_F} + {\tilde \sigma}_{xx}\right)}{\left(2/{\pi\alpha_F} + {\tilde \sigma}_{xx}\right)\left(2/{\pi\alpha_F} + {\tilde \sigma}_{yy}\right)
-\tilde\sigma_{xy}^2},
\ee
where $\alpha_F = e^2/{4\pi c\varepsilon_0\hbar}~(\ll 1)$ is the fine structure constant. Similarly we have 
\be\label{rsp}
r_{sp} = -\frac{2}{\pi\alpha_F}\frac{{\tilde \sigma}_{xy}}{\left(2/{\pi\alpha_F} + {\tilde \sigma}_{xx}\right)\left(2/{\pi\alpha_F} + {\tilde \sigma}_{yy}\right)
-{\tilde \sigma}_{xy}^2}.
\ee

Now given the fact that for 2D materials, typically we have ${\tilde \sigma}_{ij}$ to be of the order of 1, we can do an expansion in $\alpha_F$ to simplify. 
Doing this expansion of the reflection and transmission coefficient upto first order in $\alpha_F$, we obtain the complex coefficients: 
\be\label{tss_alpha}
t_{ss} \approx 1-\frac{\pi\alpha_F}{2}\frac{\sigma_{xx}}{\sigma_0}~,
\ee
\be\label{tpp_alpha}
t_{pp} \approx 1-\frac{\pi\alpha_F}{2}\frac{\sigma_{yy}}{\sigma_0}~,
\ee
and,
\be\label{rsp_alpha}
r_{sp} \approx -\frac{\pi\alpha_F}{2}\frac{\sigma_{xy}}{\sigma_0}~.
\ee

Using Eqs. \eqref{tss_alpha}-\eqref{rsp_alpha} in Eq. \eqref{Kerr} and Eq. \eqref{Faraday}, we have (upto order $\alpha_F$)
\be \label{chi_Kerr}
\chi_{\rm Kerr}^s \approx \frac{\sigma_{xy}}{\sigma_{xx}}, ~~~~{\rm and}~~~~ \chi_{\rm Kerr}^p \approx -\frac{\sigma_{xy}}{\sigma_{yy}}~,
\ee
and,
\be \label{chi_F}
\chi_{\rm Faraday}^s \approx \frac{\pi\alpha_F}{2}\sigma_{xy}, ~~~~{\rm and}~~~~ \chi_{\rm Faraday}^p \approx -\frac{\pi\alpha_F}{2}\sigma_{xy}~.
\ee
Note that $\chi_{\rm Faraday}^{s/p}$ is in general smaller by $\chi_{\rm Kerr}^{s/p}$ by a factor of $\alpha_F \approx 1/137$. Thus the Kerr angle will 
always be much larger than the Faraday angle for any 2D system in general. 

Using the simplified expression for $\chi^{s/p}_{\rm M}$, the s- component of the Kerr angle can be approximated as,
\be\label{phi_Kerr}
\tan (2\Phi^s_{\rm Kerr}) \approx \frac{2}{|\sigma_{xx}|^2}{\rm Re}\left[\sigma_{xy}(\sigma_{xx})^*\right]~.
\ee
The corresponding Kerr rotation angle for the p- component can be obtained by replacing $\sigma_{xx}$ with $-\sigma_{yy}$ in the above equation.
The Faraday angle for both s- and p- components are given by,
\be\label{phi_Faraday}
\tan 2\Phi^{s/p}_{\rm Faraday} \approx \pm \pi\alpha_F~{\rm Re}(\sigma_{xy})~.
\ee

Finally, the approximate form (lowest order in $\alpha_F$) of the ellipticity for $s$- component of the reflected beam is,
\be\label{phi_s_Kerr}
\varepsilon^s_{\rm Kerr} \approx \tan\left(\frac{1}{2}\sin^{-1}\left[\frac{2}{|\sigma_{xx}|^2} ~{\rm Im}\left[\sigma_{xy}(\sigma_{xx})^*\right]\right]\right)~.
\ee
From Eq.~\eqref{chi_Kerr}, it follows that ellipticity of the $p$- component can be obtained by replacing $\sigma_{xx}$ with $-\sigma_{yy}$ in Eq.~\eqref{phi_s_Kerr}, 
and it is given by 
\be\label{phi_p_Kerr}
\varepsilon^p_{\rm Kerr} \approx \tan\left(\frac{1}{2}\sin^{-1}\left[\frac{2}{|\sigma_{yy}|^2} ~{\rm Im}\left[\sigma_{xy}(-\sigma_{yy})^*\right]\right]\right)~.
\ee
\bibliography{refs_borophene.bib}
\end{document}